\documentclass[aps,pre]{revtex4}
\usepackage{amsmath,amssymb}
\usepackage{graphicx}
\include{epsf}
\usepackage{clrscode}

\def\(({\left(}
\def\)){\right)}
\def\[[{\left[}
\def\]]{\right]}

\newcommand{\be}{\begin{equation}}
\newcommand{\ee}{\end{equation}}
\newcommand{\bea}{\begin{eqnarray}}
\newcommand{\eea}{\end{eqnarray}}

\newcommand{\cin}{c_ {\rm in}}
\newcommand{\cout}{c_ {\rm out}}
\newcommand{\pin}{p_{\rm in}}
\newcommand{\pout}{p_{\rm out}}

\begin{document}
\title{Asymptotic analysis of the stochastic block model for modular networks and its algorithmic applications}

\author{Aurelien Decelle$^{1}$, Florent Krzakala$^{2}$, Cristopher Moore$^{3}$, Lenka Zdeborov\'a$^{4,}$}

\email[Corresponding author: ]{lenka.zdeborova@cea.fr}
\affiliation{
$^1$Universit\'e Paris-Sud \& CNRS, LPTMS, UMR8626,  B\^{a}t.~100, Universit\'e Paris-Sud 91405 Orsay, France.
$^2$CNRS and ESPCI ParisTech, 10 rue Vauquelin, UMR 7083 Gulliver, Paris 75000, France.
$^3$Santa Fe Institute and University of New Mexico, Albuquerque, New Mexico.  
$^4$Institut de Physique Th\'eorique, IPhT, CEA Saclay, and URA 2306, CNRS, 91191 Gif-sur-Yvette, France.
}

\begin{abstract}
In this paper we extend our previous work 
on the stochastic block model, a commonly used generative model for social and biological networks, and the problem of inferring functional groups or communities from the topology of the network. We use the cavity method of statistical physics to obtain an asymptotically exact analysis of the phase diagram. We describe in detail properties of the detectability/undetectability phase transition and the easy/hard phase transition for the community detection problem. Our analysis translates naturally into a belief propagation algorithm for inferring the group memberships of the nodes in an optimal way, i.e., that maximizes the overlap with the underlying group memberships, and learning the underlying parameters of the block model.  Finally, we apply the algorithm to two examples of real-world networks and discuss its performance. 
\end{abstract}

\pacs{\vspace{-1mm} 64.60.aq,89.75.Hc,75.10.Hk}
\date{\today}
\maketitle

\tableofcontents

\newpage

\section{Introduction}

Many systems of interest consist of a large number of nodes (e.g. atoms, agents, items, Boolean variables) and sometimes the only information we can observe about the system are connections, or edges, between pairs of nodes. The resulting structure is usually called a network. A naturally arising question is whether we are able to understand something about the underlying system
 based purely on the topology of the network. In many situations different nodes have different functions. For instance each of the nodes may belong to one of $q$ groups, often called functional modules or communities, and the structure of the network may depend in an \emph{a priori} unknown way on the group memberships. In general we are interested in finding how the network's function and topology
  affect each other. Hence an interesting and practically important question is whether, based on the known structure of the network, it is possible to learn in what way the group memberships influenced the structure of the network, and which node belongs to which group. 

One well-studied example of the above setting is the so-called community detection problem, for a review see e.g.~\cite{Fortunato10}. In the study of complex networks, a network is said to have community structure if it divides naturally into groups of nodes with denser connections within groups and sparser connections between groups.  This type of structure, where nodes are more likely to connect to others of the same type as in a ferromagnet, is called \emph{assortative}.  The goal is to detect the communities and to estimate, for instance, the expected number of connections within and outside of the community.  

In other cases the network can be \emph{disassortative}, with denser connections between different groups than within groups.  For instance, a set of predators might form a functional group in a food web, not because they eat each other, but because they eat similar prey.  In networks of word adjacencies in English text, nouns often follow adjectives, but seldom follow other nouns.  Even some social networks have disassortative structure: for example, some human societies are divided into moieties, and only allow marriages between different moieties.

In research on networks, generative models of random graphs often provide useful playgrounds for theoretical ideas and testing of algorithms.  The simplest such model is the Erd\H{o}s-R\'enyi random graph~\cite{ErdosRenyi59}, where every pair of nodes is connected independently with the same probability.  In this paper we study in detail the most commonly used generative model for random modular networks, the \emph{stochastic block model}, which generalizes the Erd\H{o}s-R\'enyi random graph by giving each pair of nodes a connection probability depending on what groups they belong to. We extend our previous analysis from~\cite{DecelleKrzakala11}, and provide a rather complete and asymptotically exact analysis of the phase diagram of this model. We focus on the question of whether we can infer the original group assignment based on the topology of the resulting network, and learn the unknown parameters that were used for generating the network.  We use the cavity method developed in the statistical physics of spin glasses~\cite{MezardMontanari07} to evaluate the phase diagram. 

Our results naturally translate into a message-passing algorithm called \emph{belief propagation}~\cite{YedidiaFreeman03}, that we suggest as a heuristic tool for learning parameters and inferring modules in real networks. Our results are exact in the thermodynamic limit, but the algorithm works well even on small networks as long as they are well-described by the generative model. Our theory and algorithms are straightforwardly generalizable to other generative models, including hierarchical module structures~\cite{ClausetMoore08}, overlapping modules~\cite{airoldi}, or degree-corrected versions of the stochastic block model~\cite{KarrerNewman10}.  In the present work we focus on graphs that do not contain any multiple edges or self-loops, but the generalization would be straightforward.

The paper is organized as follows.  In Section~\ref{model_def} we  define the stochastic block model.  In~\ref{bayes_inference} and~\ref{bayes_learning} we review the Bayesian theory for optimal inference and learning in the present context.  In Section~\ref{sec_works} we discuss in more detail the relation between our work and previous work on community detection, and summarize the advantages of our approach.  
In Section~\ref{bp_dep} we present the cavity method for asymptotic analysis of the model, and the associated belief propagation algorithm for parameter learning and inference of the group assignment.  In Section~\ref{fact_anal} we analyze the phase diagram and describe in detail the different phase transitions introduced in~\cite{DecelleKrzakala11}.  Finally, in Section~\ref{sec_real} we discuss applications of our approach to real-world networks.

\section{The stochastic block model}

\subsection{Definition of the model}
\label{model_def}

The stochastic block model is defined as follows.  It has parameters $q$ (the number of groups), $\{n_a\}$ (the expected fraction of nodes in each group $a$, for $1 \le a \le q$), and a $q \times q$ \emph{affinity matrix} $p_{ab}$ (the probability of an edge between group $a$ and group $b$).  We generate a random directed graph $G$ on $N$ nodes, with adjacency matrix $A_{ij}=1$ if there is an edge from $i$ to $j$ and $0$ otherwise, as follows.  Each node $i$ has a label $t_i \in \{1,\ldots,q\}$, indicating which group it belongs to.  These labels are chosen independently, where for each node $i$ the probability that $t_i=a$ is $n_a$.  Between each pair of nodes $i, j$, we then include an edge from $i$ to $j$ with probability $p_{t_i,t_j}$, setting $A_{ij}=1$, and set $A_{ij}=0$ with probability $1-p_{t_i,t_j}$.  We forbid self-loops, so $A_{ii}=0$.

We let $N_a$ denote the number of nodes in each group $a$.  Since $N_a$ is binomially distributed, in the limit of large $N$ we have $N_a/N=n_a$ with high probability.  The average number of edges from group $a$ to group $b$ is then $M_{ab} = p_{ab} N_a N_b$, or $M_{aa} = p_{aa} N_a (N_a-1)$ if $a=b$.  Since we are interested in sparse graphs where $p_{ab} = O(1/N)$, we will often work with a rescaled affinity matrix $c_{ab} = N p_{ab}$.  In the limit of large $N$, the average degree of the network is then 
\be 
c = \sum_{a,b} c_{ab} n_a n_b \, . \label{av_degree}
\ee

We can also consider the undirected case, where $A_{ij}$, $p_{ab}$, and $c_{ab}$ are symmetric.  The average degree is then
\be 
c = \sum_{a < b} c_{ab} n_a n_b + \sum_a c_{aa} \frac{n_a^2}{2} \, . \label{av_degree-directed}
\ee

Several special cases of this stochastic block model are well known and studied in the literature. Without the aim of being exhaustive let us mention three of them. We stress, however, that from a mathematical point of view many of our results are non-rigorous and hence rigorous proofs of our conjectures would be a natural and important extension of our work. 
\begin{itemize}
\item A common benchmark for community detection is the ``four groups'' test of Newman and Girvan~\cite{NewmanGirvan04}, in which a network is divided into four equally-sized groups: $n_a=1/4$, the average degree is $c=16$, and 
\be
c_{ab} = \begin{cases} c_\mathrm{in} & a=b \\ c_\mathrm{out} & a \ne b \, , \end{cases} 
\ee
where $c_\mathrm{in} > c_\mathrm{out}$ so that the community structure is assortative.  By varying the difference between $c_\mathrm{in}$ and $c_\mathrm{out}$, we create more or less challenging structures for community detection algorithms.  It is usually expected in the literature that as $N\to \infty$ it is possible to find a configuration correlated with the original assignment of nodes to the four groups as soon as the average in-degree (number of edges going to the same group) is larger than $4$, or $c/q$ in general, and that a failure to do so is due to an imperfection of the algorithm.  In contrast, from our results (see e.g. Fig.~\ref{fig_1}) it follows that in the limit $N\to \infty$, unless the average in-degree is larger than $7$ no efficient algorithm will be better than a random choice in recovering the original assignment on large networks.  At the same time we design an algorithm that finds the most likely assignment for each node if the in-degree is larger than $7$. 

\item Planted graph partitioning, a generalization of the above example, is well known in the mathematics and computer science literature.  We have $n_a=1/q$, and $p_{ab} = \pin$ if $a=b$ and $\pout$ if $a \ne b$.  We again assume an assortative structure, so that $\pin > \pout$, but now we allow the dense case where $\pin$, $\pout$ might not be $O(1/N)$. A classical result~\cite{DyerFrieze89} shows that for $\pin-\pout > O(\log{N}/N)$ the planted partition is with high probability equal to the best possible partition, in terms of minimizing the number of edges between groups. Another classical result~\cite{CondonKarp01} shows that the planted partition can be easily found as long as $\pin-\pout > O(N^{-1/2+\epsilon})$ for arbitrarily small $\epsilon$. In this work we do not aim at finding the best possible partition nor the planted partition exactly.  Rather, we are interested in conditions under which polynomial-time algorithms can find a partition that is correlated with the planted partition. In Section~\ref{sec_stab} we will show that this is possible when $\pin-\pout > \sqrt{q \pin + q(q-1)\pout} /\sqrt{N}$.  We will also argue that, depending on the values of the parameters, this task may be impossible, or exponentially hard, if this condition is not satisfied.

\item In the planted coloring problem we again have $n_a=1/q$ and $p_{ab} = \pout$ for $a \ne b$, but $p_{aa}=0$ so that there are no edges between nodes in the same group. The average degree of the graph is then $c = (q-1)\pout N/q$.  A rigorous analysis of the problem~\cite{KrivelevichVilenchik06} shows that for $c>O(q^2)$ it is possible to find in polynomial time a proper coloring correlated strongly with the planted coloring. The cavity method result derived in~\cite{KrzakalaZdeborova09} shows that configurations correlated with the planted coloring are possible to find if and only if $c > c_q$, where the critical values $c_q$ are given in~\cite{KrzakalaZdeborova09}.  For large $q$ we have $c_q = O(2q\log{q})$. However, it is known how to find such colorings in polynomial time only for $c>(q-1)^2$~\cite{KrzakalaZdeborova09}, and there are physical reasons to believe that $c=(q-1)^2$ provides a threshold for success of a large class of algorithms, as we will explain later.
\end{itemize}

We stress that in this paper we are mostly interested in the large $N$ behavior of the generative model and hence in the text of the paper we will neglect terms that are negligible in this limit. For instance, we will write that the number of edges is $M=cN/2$ even if for finite size $N$ one typically has fluctuations around the average of size $O(\sqrt{N})$.

Now assume that a network was generated following the stochastic block model described above. The resulting graph $G$ is known, but the parameters $q$, $n_a$, $p_{ab}$ and the labeling $t_i$ are unknown. In this paper we address the two following questions:
\begin{itemize}
   \item[(i)] Given the graph $G$, what are the most likely values of the parameters $q$, $n_a$, $p_{ab}$ that were used to generate the graph? We will refer to this question as \emph{parameter learning}. 
   \item[(ii)] Given the graph $G$ and the parameters $q$, $n_a$, $p_{ab}$, what is the most likely assignment of a label (group) to a given node? 
In particular, is this most likely assignment better than a random guess? How much better? We will refer to this to as \emph{inferring the group assignment}. 
\end{itemize}

In order to be able to give a quantitative answer to the second question we define \emph{agreement} between the original assignment $\{t_i\}$ and its estimate $\{q_i\}$ as
\be
       A(\{t_i\},\{q_i\}) = \max_{\pi} \frac{1}{N}\sum_{i} \delta_{t_i,\pi(q_i)}   \, , \label{agreement}
\ee
where $\pi$ ranges over the permutations on $q$ elements.  We also define a normalized agreement that we call the \emph{overlap},
\be
       Q(\{t_i\},\{q_i\}) = \max_{\pi}\frac{ \frac{1}{N}\sum_{i} \delta_{t_i,\pi(q_i)} - \max_a n_a }{1-\max_a n_a}  \, . \label{overlap}
\ee
The overlap is defined so that if $t_i=q_i$ for all $i$, i.e., if we find the exact labeling, then $Q=1$. If on the other hand the only information we have are the group sizes $n_a$, and we assign each node to the largest group to maximize the probability of the correct assignment of each node, then $Q=0$. We will say that a labeling $\{q_i\}$ is correlated with the original one $\{t_i\}$ if in the thermodynamic limit $N\to \infty$ the overlap is strictly positive, with $Q > 0$ bounded above some constant.  

Our main results provide exact answers to the above questions in the thermodynamic limit for sparse networks, i.e. when $c_{ab}=N p_{ab} = O(1)$ and $n_a=O(1)$ are constants independent of size as $N \to \infty$.  Many real-world networks are sparse in that the total number of edges grows only as $O(N)$ rather than $O(N^2)$ (although we do not address networks with heavy-tailed degree distributions).  In the dense case where the average degree diverges as $N\to \infty$, learning and inference are algorithmically easier, as was previously realized in~\cite{CondonKarp01,BickelChen09} and as will also become clear from the large-degree limit of our results.

\subsection{Optimal inference of the group assignment}
\label{bayes_inference}

The probability that the stochastic block model generates a graph $G$, with adjacency matrix $A$, along with a given group assignment $\{q_i\}$, conditioned on the parameters $\theta = \{ q,\{n_a\},\{p_{ab}\}\}$ is
\be
    P(G,\{q_i\} \!\mid\! \theta ) = \prod_{i \ne j} \left[  p_{q_i,q_j}^{A_{ij}} (1-p_{q_i,q_j})^{1-A_{ij}}\right] \prod_i n_{q_i} \, , 
    \label{Pgraph} 
\ee
where in the undirected case the product is over pairs $i < j$.  Note that the above probability is normalized, i.e. $\sum_{G,\{q_i\}} P(G,\{q_i\} \!\mid\! \theta )=1$. Assume now that we know the graph $G$ and the parameters $\theta$, and we are interested in the probability distribution over the group assignments given that knowledge. Using Bayes' rule we have
\be
    P(\{q_i \} \!\mid\! G ,\theta ) = \frac{P(G,\{q_i\} \!\mid\! \theta )}{\sum_{t_i}P(G, \{t_i\} \!\mid\! \theta ) } \, .
\ee
In the language of statistical physics, this distribution is the Boltzmann distribution of a generalized Potts model with Hamiltonian
\be
    H(\{q_i\} \!\mid\! G, \theta) =- \sum_i \log{n_{q_i}}-\sum_{i \ne j} \left[ {A_{ij} \log{ c_{q_i,q_j}}} + {(1-A_{ij}) \log{ \left(1-\frac{c_{q_i,q_j}}{N}\right)}}\right] \, , \label{Ham}
\ee
where the sum is over pairs $i < j$ in the undirected case.  The labels $q_i$ are Potts spins taking one of the $q$ possible values, and the group sizes $n_{q_i}$ (or rather their logarithms) become local magnetic fields.  In the sparse case $c_{ab}=O(1)$, there are strong $O(1)$ interactions between connected nodes, $A_{ij}=1$, and weak $O(1/N)$ interactions between nodes that are not connected, $A_{ij}=0$. Recall that the Boltzmann distribution at unit temperature is
\be
    \mu(\{q_i\} \!\mid\! G, \theta)   
    =   P(\{q_i\} \!\mid\! G, \theta) 
    = \frac{  e^{-H(\{q_i\} \mid G, \theta)}  }{ \sum_{\{q_i\}}  e^{-H(\{q_i\} \mid G, \theta) } }\, , \label{Boltz}
\ee
where the denominator is the corresponding partition function 
\be 
     Z(G,\theta)  = \sum_{\{q_i\}}  e^{-H(\{q_i\} \mid G, \theta) } \, . \label{part}
\ee 
Note that we define $H(\{q_i\}|G,\theta) = -\log P(G,\{q_i\} \!\mid\! \theta) - M \log{N}$ to keep the formally useful property that the energy is extensive, i.e., proportional to $N$, in the thermodynamic limit.  In statistical physics it is usual to work with the free energy density 
\be
    \frac{F_N(G,\theta)}{N} = -\frac{\log{Z(G,\theta)}}{N} \underset{N \to \infty}{\rightarrow} f(G, \theta )\, . \label{Fnrg}
\ee
Since the energy (\ref{Ham}) is extensive, the free energy density has a well defined finite thermodynamic limit $f(G,\theta)$.

It is useful to notice that if the parameters $q,\{n_a\},\{c_{ab}\}$ are known then the Boltzmann distribution (\ref{Boltz}) is asymptotically uniform over all configurations with the right group sizes and the right number of edges between each pair of groups, $N_a/N = n_a$ and $M_{ab}/N = c_{ab} n_a n_b$. The original, correct group assignment is just one of these configurations.  In a statistical physics sense, the original group assignment is an equilibrium configuration for the Boltzmann distribution, rather than its ground state.  In particular, if we were presented with the original assignment $\{t_i\}$ and a typical assignment $\{q_i\}$ sampled according to the Boltzmann distribution, we would be unable to tell which one was correct.

The marginals of the Boltzmann distribution, i.e. the probabilities $\nu_i(q_i)$ that a node $i$ belongs to a group $q_i$, are
\be
        \nu_i(q_i) = \sum_{\{q_j\}_{j\neq i}}  \mu(\{q_j\}_{j\neq i},q_i)\, . \label{nu_i}
\ee
Our estimate $q_i^*$ of the original group assignment assigns each node to its most-likely group, 
\be
              q_i^* = \mathrm{argmax}_{q_i} \nu_i(q_i) \, . \label{marginalization}
\ee
If the maximum of $\nu_i(q_i)$ is not unique, we choose at random from all the $q_i$ achieving the maximum.  We refer to this method of estimating the groups as \emph{marginalization}; in Bayesian inference $q_i^*$ is called the \emph{maximum posterior marginal}.  Standard results in Bayesian inference (e.g.~\cite{Iba99}) show that it is in fact the optimal estimator of the original group assignment $\{t_i\}$ if we seek to maximize the number of nodes at which $t_i=q^*_i$.  In particular, the ground state of the Hamiltonian~\eqref{Ham}, i.e. the configuration $\{q_i^{\rm gs}\}$ that maximizes $\mu(\{q_i\})$, has in general a slightly smaller overlap with the original assignment than $\{q_i^*\}$ does.

Note that the Boltzmann distribution is symmetric with respect to permutations of the group labels.  Thus the marginals over the entire Boltzmann distribution are uniform.  However, if this permutation symmetry is broken in the thermodynamic limit, so that each permutation corresponds to a different Gibbs state, we claim that marginalization within one of these Gibbs states is the optimal estimator for the overlap defined in~\eqref{overlap}, where we maximize over all permutations.

One of the advantages of knowing the marginals of the Boltzmann distribution~\eqref{Boltz} is that in the thermodynamic limit we can evaluate the overlap $Q(\{t_i\},\{q_i^*\})$ even without the explicit knowledge of the original assignment $\{t_i\}$. It holds that 
\be
        Q_{\rm margin}  \equiv \lim_{N\to \infty} \frac{\frac{1}{N} \sum_i \nu_i(q^*_i) - \max_a n_a}{1-\max_a n_a} =  \lim_{N\to \infty} Q(\{t_i\},\{q_i^*\})   \, . \label{over_best}
\ee 
The overlap $Q_{\rm margin}$ measures the amount of information about the original assignment that can be retrieved from the topology of the network, given the parameters $\theta$.  The marginals $\nu_i(q_i)$ can also be used to distinguish nodes that have a very strong group preference from those that are uncertain about their membership. 

Another interesting property is that two random configurations taken from the Boltzmann distribution (\ref{Boltz}) have the same agreement as a random configuration with the original assignment $t_i$, i.e.
\be
  \lim_{N\to \infty} \frac{1}{N}\max_{\pi} \sum_{i} \nu_i\left( \pi(t_i) \right) = \lim_{N\to \infty} \frac{1}{N} \sum_i \sum_a \nu_i(a)^2 \, , \label{m=q}
\ee
where $\pi$ again ranges over the permutations on $q$ elements.  This identity holds only if the associated Boltzmann distribution was computed using the correct parameters $\theta$. In the statistical physics of spin glasses, the property (\ref{m=q}) is known as the equality between the Edwards-Anderson overlap and the magnetization, and holds on the Nishimori line.  Here the knowledge of the actual parameter values $\theta$ is equivalent to the Nishimori condition being satisfied; for more details see Section~\ref{bayes_learning}.

\subsection{Learning the parameters of the model}
\label{bayes_learning}

Now assume that the only knowledge we have about the system is the graph $G$.  The general goal in machine learning is to learn the most probable values of the parameters $\theta$ of an underlying model based on the data known to us.  In this case, the parameters are $\theta=\{ q,\{n_a\},\{c_{ab}\} \}$ and the data is the graph $G$, or rather the adjacency matrix $A_{ij}$.  According to Bayes' rule, the probability $P(\theta \!\mid\! G)$ that the parameters take a certain value, conditioned on $G$, is proportional to the probability $P(G \!\mid\! \theta)$ that the model with parameters $\theta$ would generate $G$.  This in turn is the sum of $P(G,\{q_i\} \!\mid\! \theta)$ over all group assignments $\{q_i\}$:
\be
   P(\theta \!\mid\! G)  = \frac{P(\theta)}{P(G)} \,P(G \!\mid\! \theta) = \frac{P(\theta)}{P(G)} \sum_{\{q_i\}}  P(G,\{q_i\} \!\mid\! \theta) \, . \label{P_theta}
\ee

In Bayesian inference, $P(\theta \mid G)$ is called the \emph{posterior distribution}.  The \emph{prior distribution} $P(\theta)$ includes any graph-independent information we might have about the values of the parameters.  In our setting, we wish to remain perfectly agnostic about these parameters; for instance, we do not want to bias our inference process towards assortative structures.  Thus we assume a uniform prior, i.e., $P(\theta)=1$ up to normalization.  Note, however, that since the sum in (\ref{P_theta}) typically grows exponentially with $N$, we could take any smooth prior $P(\theta)$ as long as it is independent of $N$; for large $N$, the data would cause the prior to ``wash out,'' leaving us with the same posterior distribution we would have if the prior were uniform.

Thus maximizing $P(\theta \mid G)$ over $\theta$ is equivalent to maximizing the partition function (\ref{part}) over $\theta$, or equivalently minimizing the free energy density defined in Eq.~(\ref{Fnrg}) of the Potts model (\ref{Ham}) as a function of $\theta$.  As in the saddle-point method, if the function $f(\theta)$ has a non-degenerate minimum, then in the thermodynamic limit this minimum is achieved with high probability at precisely the values of the parameters that were used to generate the network. In mathematical terms, if we call $\theta^*$ the original parameters and $\tilde{\theta}$ the ones minimizing $f(\theta)$, then for all $\epsilon>0$ the probability $\lim_{N\to \infty} \Pr[ |\theta^*-\tilde{\theta}|<\epsilon] = 1$. It is also important to note that due to the self-averaging nature of the model, as $N \to \infty$ the free energy depends only on $\theta$ and not on the precise realization of the network.  
We will study the free energy in detail in the next section, but for the moment suppose that it indeed has a non-degenerate minimum as a function of $\theta$.  In that case we can learn the exact parameters: the number of groups $q$, their sizes $\{n_a\}$, and the affinity matrix $\{c_{ab}\}$.  

In some cases, rather than minimizing $f(\theta)$ directly, it is useful to write explicit conditions for the stationarity of $f(\theta)$.  Taking the derivative of $f(\theta)$ with respect to $n_a$ for $1 \le a \le q$, subject to the condition $\sum_a n_a=1$, and setting these derivatives equal to zero gives 
\be
    \frac{1}{N} \sum_i  \langle \delta_{q_i,a} \rangle = \frac{\langle N_a \rangle}{N} = n_a  \quad \forall a=1,\ldots,q \, , \label{na_Nish}
\ee
where by $\langle f(\{q_i\}) \rangle = \sum_{\{q_i\}} f(\{q_i\}) \mu(\{q_i\}|G,\theta)$ we denote the thermodynamic average.  Thus for each group $a$, the most likely value of $n_a$ is the average group size; an intuitive result, but one that deserves to be stated.  Analogously, taking the derivative of $f(\theta)$ by the affinities $c_{ab}$ gives
\be
     \frac{1}{N n_a n_b} \sum_{(i,j) \in E}  \langle \delta_{q_i,a} \delta_{q_j,b} \rangle 
     = \frac{\langle M_{ab} \rangle}{N n_a n_b}
     = c_{ab}  \quad \forall a, b \, . \label{ab_Nish} 
\ee
Meaning that the most likely value of $c_{ab}$ is proportional to the average number of edges from group $a$ to group $b$.  More to the point, the most likely value of $p_{ab}=c_{ab}/N$ is the average fraction of the $N_a N_b$ potential edges from group $a$ to group $b$ that in fact exist. In the undirected case, for $a=b$ we have 
\be
 \frac{1}{N n_a^2/2} \sum_{(i,j) \in E} \langle \delta_{q_i,a} \delta_{q_j,a} \rangle 
 = \frac{\langle M_{aa} \rangle}{N n_a^2 / 2} = c_{aa} \quad \forall a \, . \label{aa_Nish}
\ee 

The stationarity conditions (\ref{na_Nish}--\ref{aa_Nish}) naturally suggest an iterative way to search for the parameters $\theta$ that minimize the free energy.  We start with arbitrary estimates of $\theta$ (actually not completely arbitrary, for a more precise statement see subsequent sections), measure the mean values $\langle N_a \rangle$ and $\langle M_{ab} \rangle$ in the Boltzmann distribution with parameters $\theta$, and update $\theta$ according to (\ref{na_Nish}--\ref{aa_Nish}) .  We then use the resulting $\theta$ to define a new Boltzmann distribution, again measure $\langle N_a \rangle$ and $\langle M_{ab} \rangle$, and so on until a fixed point is reached.  

In statistical physics, the stationarity conditions (\ref{na_Nish}--\ref{aa_Nish}) can be interpreted as the equality of the quenched and annealed magnetization and correlations.  In models of spin glasses (e.g.~\cite{Iba99}) they are sometimes referred to as the \emph{Nishimori conditions}. This iterative way of looking for a maximum of the free energy is equivalent to the well-known \emph{expectation-maximization} (EM) method in statistics~\cite{DempsterLaird77}.
 
Note that if the conditions (\ref{na_Nish}--\ref{aa_Nish}) are satisfied, the average of the Hamiltonian in the Boltzmann distribution (\ref{Boltz}) can be easily computed as 
\be
    \frac{1}{N} \langle H \rangle \equiv e = -\sum_a n_a \log{n_a} - \sum_{a,b} n_a n_b c_{ab} \log{c_{ab}} 
    + c \, , \label{dir-energy}
\ee
where the term $c$ comes from the weak $O(1/N)$ interactions between disconnected pairs of nodes.  
In the undirected case, 
\be
    \frac{1}{N} \langle H \rangle \equiv e 
    = -\sum_a n_a \log{n_a} - \sum_{a < b} n_a n_b c_{ab} \log{c_{ab}} 
    - \sum_a \frac{n_a^2}{2} c_{aa} \log{c_{aa}} 
    + \frac{c}{2} \, . \label{energy}
\ee

Note that as usual in statistical physics the free energy is the average of the energy minus the entropy, $f=e-s$ (where here the ``temperature'' is unity). The entropy is the logarithm of the number of assignments that have the corresponding energy. In the Bayesian inference interpretation the entropy counts assignments that are as good as the original one (i.e. they have the same group sizes and the same number of edges between each pair of groups). This entropy provides another useful measure of significance of communities in the network, and was studied numerically for instance in~\cite{GoodMontjoye10}.

\subsection{Our contribution and relation to previous work}
\label{sec_works}

The Bayesian approach presented in the previous two sections is well known in machine learning and statistics.  However, it is also well known that computing the partition function and the marginal probabilities of a model defined by the Hamiltonian (\ref{Ham}) or computing the averages on the left-hand sides of (\ref{na_Nish}--\ref{aa_Nish}) is a computationally hard task. In general, and our model is no exception, there is no exact polynomial-time algorithm known for these problems.

A standard tool of statistical physics and statistical inference is to approximate thermodynamic averages using Monte Carlo Markov chains (MCMC), also known as Gibbs sampling. Any Markov chain respecting detailed balance will, after a sufficient number of steps, produce sample configurations according to the Boltzmann distribution~(\ref{Boltz}). This can then be used to compute averages like $\langle N_a \rangle$ and $\langle M_{ab} \rangle$, or even the free energy if we integrate over a temperature-like parameter.  A central question, however, is the equilibration time of these Markov chains.  For very large networks, a running time that grows more than linearly in $N$ is impractical.

The running time of Gibbs sampling is one reason why the Bayesian approach is not that widely used for community detection. Exceptions include specific generative models for which the partition function computation is tractable~\cite{NewmanLeicht07,BallKarrer11}, the work of~\cite{HofmanWiggins08} where a variational approximation to bound the partition function (or related expectations) was used, the work of~\cite{ClausetMoore08} where a more elaborate generative model is used to infer hierarchies of communities and subcommunities.  Another exception is~\cite{MooreKDD}, where Gibbs sampling is used to estimate the mutual information between each node and the rest of the graph in order to perform ``active learning.'' 

The main contribution of this work is a detailed, and in the thermodynamic limit exact, analysis of the stochastic block model and its phase diagram with the use of the cavity method~\cite{MezardParisi01,MezardMontanari07} developed in the theory of spin glasses.  We show that there is a region in the phase diagram, i.e., a range of parameters $\theta$, where inference is impossible even in principle, 
since the marginals of the Boltzmann distribution yield no information about the original group assignment.  There is another region where inference is possible but, we argue, exponentially hard. Finally, there is a region where a belief propagation algorithm~\cite{YedidiaFreeman03}, that emerges naturally from the cavity method, computes the free energy density and corresponding expectations exactly in the thermodynamic limit, letting us infer the parameters optimally and in linear time.  As we show later, this algorithm also performs very well for networks of moderate size, and it is useful for real-world networks as well.

The boundaries between these phases correspond to well-known phase transitions in the statistical physics of spin glasses: namely, the dynamical transition or the reconstruction threshold, see e.g.~\cite{MezardMontanari06,ZdeborovaKrzakala07}; the condensation transition or the Kauzmann temperature~\cite{KrzakalaMontanari06,Kauzmann48}; and the easy/hard transition in planted models introduced in~\cite{KrzakalaZdeborova09}. There is also a close relation between our approach and the optimal finite-temperature decoding~\cite{Nishimori93,Sourlas94,Iba99,NishimoriBook01}, and the statistical mechanics approach to image processing~\cite{Tanaka02}. 

In fact, the theory of spin glasses also leads to the conclusion that Gibbs sampling works in linear time, i.e., its equilibration time is linear in $N$, in the same region where belief propagation works. However, belief propagation is considerably faster than Gibbs sampling at finding the marginals, since belief propagation produces marginals directly while Gibbs sampling requires us to measure them by taking many independent samples.

Belief propagation was previously suggested as an algorithm for detecting communities in~\cite{Hastings06}.  However, in that work its performance was not studied systematically as a function of the parameters of the block model.  Moreover, there the parameters $\theta$ were fixed to an assortative community structure similar to the planted partition model discussed above, rather than being learned.

We argue that our Bayesian approach, coupled with belief propagation to compute marginals and estimate the free energy, is optimal for graphs generated from the stochastic block model.  For such random networks it possesses several crucial advantages over the methods that are widely used for community detection in the current literature, without being computationally more involved.  For a review of methods and results known about community detection see for instance~\cite{Fortunato10} and references therein. Let us list some of the advantages: 
\begin{itemize}
\item{\bf General modules, not just assortative ones:} It is fair to say that when there are well-separated assortative communities in the network, the community detection problem is relatively easy, and many efficient algorithms are available in the literature; see for instance~\cite{LancichinettiFortunato09} for a comparative study.  However, for block models where the matrix of affinities $p_{ab}$ is not diagonally dominant, i.e., where functional groups may be disassortative or where edges between them are directed, the spectrum of available algorithms is, so far, rather limited.
\item {\bf No prior knowledge of parameters needed:}
Our method does not require any prior knowledge about the functional groups or how they tend to be connected.  It is able to learn the number of groups $q$, their sizes $n_a$, and the affinity matrix $c_{ab}$.
\item{\bf Asymptotically exact for the stochastic block model:}
Unlike any other known method, for networks that are created by the stochastic block model, in the limit $N \to \infty$ our algorithm either outputs the exact parameters $q,n_a,c_{ab}$, or halts and reports that learning is impossible or algorithmically hard.  In the second case we are very confident that no other method will be able to do better in terms of the overlap parameter (\ref{overlap}), and we will explain the reasons for this conjecture later in the text.
\item{\bf Detects when a network has no communities:}   While people may differ on the precise definition of community structure and how to find it, they presumably agree that a purely random graph, such as an Erd\H{o}s-R\'enyi graph where all pairs of nodes are connected with the same probability, has no community structure to discover.  However, the vast majority of community detection algorithms do in fact find illusory communities in such graphs, due to random fluctuations.  For instance, sparse random graphs possess bisections, i.e., divisions of the nodes into two equal groups, where the number of edges between groups is much smaller than the number of edges within groups. To give a specific example, nodes in a large 3-regular random graph can be bisected in such a way that only about $11\%$ of all edges are between the groups~\cite{SulcZdeborova10}.  Popular measures of community significance such as modularity~\cite{NewmanGirvan04} do not take this fact into account.  In contrast, our method naturally recognizes when a network does not in fact contain any modular structure.
\item {\bf Better measures of significance:}
More generally, most known methods for community detection aim at providing one assignment of nodes to groups; physically speaking, they look for a single ground state.  However, there are usually a large number of group assignments that are comparable to each other according to various quality measure, see e.g.~\cite{GoodMontjoye10}.  Our method provides all the thermodynamically significant group assignments ``at once,'' by providing the marginal probabilities with which each node belongs to a given community.  It also provides the entropy of the good assignments, giving us a measure of how non-unique they are.  This kind of information is much more informative than a single group assignment, even if we can find the ``best'' one. Physically speaking, the Boltzmann distribution tells us more about a network than the ground state does.
\end{itemize}

We stress, however, that all of the above is true only for networks generated from the stochastic block model, or for real world networks that are well described by this generative model.  Indeed, many of the other methods suggested in the literature for community detection also implicitly assume that the network under study is well described by a similar model.  For many real networks, this is not true; for instance, as pointed out in~\cite{KarrerNewman10}, the block model performs poorly on networks where communities include nodes with a very broad range of degrees.  On the other hand, the Bayesian approach and belief propagation algorithm presented here can be generalized straightforwardly to any generative model where the likelihood (\ref{Pgraph}) can be written as a product of local terms, such as the degree-corrected block model suggested in~\cite{KarrerNewman10}.  Thus our methods can apply to a wide variety of generative models, which take various kinds of information about the nodes and edges into account.  We leave these generalizations for future work.

\section{Cavity method and the stochastic block model}
\label{bp_dep}

In this section we derive the cavity equations and the associated belief propagation (BP) algorithm for computing the marginal probabilities (\ref{nu_i}) and the average values (\ref{na_Nish}--\ref{aa_Nish}) we need to learn the parameters $\theta$.  When applied in the thermodynamic limit $N \to \infty$, our analysis lets us describe the phase diagram of the learning and inference problems, showing for which parameters these problems are easy, hard, or impossible.

\subsection{Cavity equations: marginals and free energy}

In the literature the so-called \emph{replica symmetric cavity method} is often understood in terms of the BP algorithm and its behavior in the thermodynamic limit with properly chosen initial conditions.  From a physics point of view, the BP algorithm is an iterative way to compute the partition function by neglecting correlations between the neighbors of node $i$ while conditioning on the ``spin'' or label of node $i$.  Such correlations are non-existent if the network of interactions is a tree.  On networks that are locally treelike, if correlations decay rapidly with distance these correlations are negligible, making BP asymptotically exact.

Note that in our case the ``network of interactions'' is fully connected, since in the Hamiltonian~\eqref{Ham} there are weak interactions even along the non-edges, i.e., between pairs of nodes that are not connected.  However, as we will see these weak interactions can be replaced with a ``mean field,'' limiting the interactions to the sparse network.

The belief propagation equations are derived from a recursive computation of the partition function with the assumption that the network of interactions is a tree.  
The asymptotic exactness of the replica symmetric cavity method (BP equations) is then validated by showing that the correlations that have been neglected are indeed negligible in the thermodynamic limit. 

To write the belief propagation equations for the Hamiltonian (\ref{Ham}) we define conditional marginals, or \emph{messages}, denoted $\psi_{q_i}^{i\to j}$.  This is the marginal probability that the node $i$ belongs to group $q_i$ in the absence of node $j$.  The cavity method assumes that the only correlations between $i$'s neighbors are mediated through $i$, so that if $i$ were missing---or if its label were fixed---the distribution of its neighbors' states would be a product distribution.  In that case, we can compute the message that $i$ sends $j$ recursively in terms of the messages that $i$ receives from its other neighbors $k$:
\be
   \psi_{t_i}^{i\to j} = \frac{1}{Z^{i\to j}} \, n_{t_i} \prod_{k\in \partial i\setminus j} \left[ \sum_{t_k}  c^{A_{ik}}_{t_i t_k} \left(1-\frac{c_{t_i t_k}}{N}\right)^{1-A_{ik}} \psi_{t_k}^{k\to i}  \right] \, , \label{BP_iter_exact}   
\ee
where $\partial i$ denotes $i$'s neighborhood, and $Z^{i\to j}$ is a normalization constant ensuring $\sum_{t_i}\psi_{t_i}^{i\to j} =1$.  
We apply~\eqref{BP_iter_exact} iteratively until we reach a fixed point $\{\psi_{q_i}^{i\to j}\}$.  Then the marginal probability is estimated to be $\nu_i(t_i)=\psi_{t_i}^{i}$, where
\be
    \psi_{t_i}^{i} = \frac{1}{Z^{i}} \, n_{t_i} \prod_{k\in \partial i} \left[ \sum_{t_k}  c^{A_{ik}}_{t_i t_k} \left(1-\frac{c_{t_i t_k}}{N}\right)^{1-A_{ik}}\psi_{t_k}^{k\to i}  \right] \, .
\ee
These equations are for the undirected case.  In a directed network, $i$ would send and receive messages from both its incoming and outgoing neighbors, and we would use $c_{t_k, t_i}$ or $c_{t_i, t_k}$ for incoming and outgoing edges respectively.

Since we have nonzero interactions between every pair of nodes, we have potentially $N(N-1)$ messages, and indeed~(\ref{BP_iter_exact}) tells us how to update all of these for finite $N$.  However, this gives an algorithm where even a single update takes $O(N^2)$ time, making it suitable only for networks of up to a few thousand nodes.  Happily, for large sparse networks, i.e., when $N$ is large and $c_{ab}=O(1)$, we can neglect terms of sub-leading order in $N$.  In that case we can assume that $i$ sends the same message to all its non-neighbors $j$, and treat these messages as an external field, so that we only need to keep track of $2M$ messages where $M$ is the number of edges.  In that case, each update step takes just $O(M)=O(N)$ time.

To see this, suppose that $(i,j)\notin E$.  We have 
\be
 \psi_{t_i}^{i\to j}  = \frac{1}{Z^{i\to j}} \, n_{t_i}   \prod_{k\notin \partial i\setminus j} \left[ 1-\frac{1}{N}\sum_{t_k}c_{t_k t_i} \psi_{t_k}^{k\to i}  \right] \prod_{k\in \partial i} \left[ \sum_{t_k}  c_{t_k t_i} \psi_{t_k}^{k\to i}  \right]   = \psi_{t_i}^{i} +O\left(\frac{1}{N}\right) \, .
\ee
Hence the messages on non-edges do not depend to leading order on the target node $j$.  On the other hand, if $(i,j)\in E$ we have 
\be
 \psi_{t_i}^{i\to j}  = \frac{1}{Z^{i\to j}} \, n_{t_i}   \prod_{k\notin \partial i} \left[ 1-\frac{1}{N}\sum_{t_k}c_{t_k t_i} \psi_{t_k}^{k\to i}  \right] \prod_{k\in \partial i\setminus j} \left[ \sum_{t_k}  c_{t_k t_i} \psi_{t_k}^{k\to i}  \right]  \, .
\ee
The belief propagation equations can hence be rewritten as
\be
\psi_{t_i}^{i\to j}   =
\frac{1}{Z^{i\to j}} \, n_{t_i} e^{-h_{t_i}} \prod_{k\in \partial i\setminus j} \left[ \sum_{t_k}  c_{t_k t_i} \psi_{t_k}^{k\to i}  \right] \, ,  \label{BP_iter} 
\ee
where we neglected terms that contribute $O(1/N)$ to $\psi^{i\to j}$, and defined an auxiliary external field 
\be
        h_{t_i} = \frac{1}{N}\sum_{k} \sum_{t_k}c_{t_k t_i} \psi_{t_k}^{k}\, . \label{ex_field}
\ee
In order to find a fixed point of Eq.~(\ref{BP_iter}) in linear time we update the messages $\psi^{i\to j}$, recompute $\psi^j$, update the field $h_{t_i}$ by adding the new contribution and subtracting the old one, and repeat.  The estimate of the marginal probability $\nu_i(t_i)$ is then
\be
   \psi_{t_i}^{i}  = \frac{1}{Z^{i}} \, n_{t_i} e^{-h_{t_i}} \prod_{j\in \partial i} \left[ \sum_{t_j}  c_{t_j t_i} \psi_{t_j}^{j\to i}  \right] \label{BP_marg} \, .
\ee
When the cavity approach is asymptotically exact then the true marginal probabilities obey $\nu_i(t_i)=\psi_{t_i}^{i}$. The overlap with the original group assignment is then computed from (\ref{over_best}). Introducing 
\bea
    Z^{ij} &=& \sum_{a < b} c_{ab} ( \psi^{i\to j}_a \psi_b^{j\to i}+ \psi^{i\to j}_b \psi_a^{j\to i}) + \sum_a c_{aa}  \psi^{i\to j}_a \psi_a^{j\to i} \quad {\rm for} \quad (i,j)\in E  \label{Z_ij} \\
    \tilde{Z}^{ij} &=&  \sum_{a,b} \left( 1-\frac{c_{ab}}{N} \right) \psi^{i}_a \psi_b^{j} \quad {\rm for} \quad (i,j) \notin E\, , \, , \\
    Z^i&=& \sum_{t_i} n_{t_i} e^{-h_{t_i}} \prod_{j\in \partial i}  \sum_{t_j}  c_{t_j t_i} \psi_{t_j}^{k\to i}  
\eea
we can write the BP estimate for the free energy, also called the \emph{Bethe free energy}, in the thermodynamic limit as
\be
    f_{\rm BP}(q,\{n_a\},\{c_{ab}\}) =  - \frac{1}{N} \sum_i \log{Z^i} + \frac{1}{N}\sum_{(i,j) \in E}  \log{Z^{ij}} - \frac{c}{2} \label{Bethe_fe}\, ,
\ee
where $c$ is the average degree given by (\ref{av_degree}) and the third term comes from the edge-contribution of non-edges (i.e. $\sum_{i,j} \log(\tilde{Z}^{ij})$). 

When deriving the BP equations (\ref{BP_iter}), the BP estimates of marginals (\ref{BP_marg}), and the Bethe free energy (\ref{Bethe_fe}), we saw that even though the original graph of interactions is fully connected we end up with BP equations on the edges of the original network.  This network is locally treelike, so standard assumptions about correlation decay suggest that the BP equations are then asymptotically exact.

In the statistical physics of spin glasses the equations presented in this section are called the \emph{replica symmetric cavity equations}.  A large amount of work has been devoted to understanding under what circumstances these equations give asymptotically exact results for the marginals and the free energy, and when they do not~\cite{MezardParisi87b,MezardMontanari07}. All known cases when the replica symmetric solution of a model like (\ref{Ham}) is not correct on a randomly generated network can be divided into two classes: a static spin glass phase with spin glass susceptibility 
\be
   \chi_{\rm SG} = \frac{1}{N} \sum_{i,j} \sum_{t_i,t_j} [   \nu_{ij}(t_i,t_j)   -\nu_i(t_i)\nu_j(t_j)  ]^2 \label{susc_SG}
\ee
diverging with the system size, or a first order phase transition into a dynamically non-attractive ``ferromagnetic'' phase, see e.g.~\cite{FranzMezard01}.

It is a general property of inference problems that at the correct value of the parameters $\theta$ the static spin glass phase never exists. Indeed, when the parameters $\theta$ are the actual ones from which the graph was generated, the system satisfies the so-called Nishimori condition, and a very general result is that there is no static spin glass phase on the Nishimori line~\cite{NishimoriBook01,Montanari08}. 

On the other hand, the first-order phase transition to a dynamically non-attractive ferromagnetic phase cannot be avoided.  This phase is easy to detect, and to describe asymptotically exactly with the BP equations, if we know the true group assignment.  Thus the BP analysis is still asymptotically exact for the purpose of analysis of the phase diagram.  However, in the situation we care about, where the original assignment is not known, this phase transition poses an algorithmic problem.  We explain this in detail in Section~\ref{fact_anal}.

\subsection{Belief propagation algorithm for inferring the group assignment}

We present here the belief propagation algorithm for inferring the group assignment, and for estimating the free energy, the marginals, and the overlap.

\begin{codebox}
\Procname{$\proc{BP-inference}(q,n_a,c_{ab},A_{ij},{\rm criterium},t_{\rm max}$)}
\li Initialize randomly $q$-component normalized vector $\{\psi^{i\to j}\}$ for each edge $(i,j)$; \label{line_init}
\li For each node $i$ compute message $\psi^i_{s_i}$ according to~(\ref{BP_marg});
\li Compute the $q$-component auxiliary field $h_{t}$ according to~(\ref{ex_field});
\li ${\rm conv} \gets {\rm criterium}+10$; $t \gets 0$;
\li \While ${\rm conv}>{\rm criterium}$ and $t<t_{\rm max}$:
\li     \Do ${\rm conv} \gets 0$; $t \gets t+1$;
\li           \For every message $\psi^{i\to j}$ (in random order):
\li    \Do  Update all $q$-components of $\psi^{i\to j}$ according to~(\ref{BP_iter}); 
\li        ${\rm conv} \gets {\rm conv} + |\psi_{\rm new}^{i\to j}-\psi_{\rm old}^{i\to j} |$
\li        Update $\psi^j$ using the new value of $\psi^{i\to j}$ and~(\ref{BP_marg});
\li        Update the field $h$ by subtracting the old $\psi^j$ and adding the new value~(\ref{ex_field});
         \End
    \End 
\li Compute free energy according to eqs. (\ref{Z_ij}--\ref{Bethe_fe});
\li \Return free energy
\li \Return messages $\{\psi^{i\to j}\}$
\li \Return group assignment $q_i^*={\rm argmax}_q \psi_q^i$
\li \Return overlap (\ref{over_best}) computed with $\nu_i=\psi^i$ 
\end{codebox}

The main cycle of the algorithm takes $O(N)$ time.  Specifically, $2M=cN$ messages need to be updated, and each such update takes $O(c)=O(1)$ time operations. The number of iterations needed for the algorithm to converge is, in general, a more complicated question. However, at the right value of the parameters $\theta$, the Nishimori condition ensures that BP will converge to a fixed point in a constant number ($t_{\rm max}$) of steps, so that the total running time of the algorithm is $O(N)$.  This is illustrated in Fig.~\ref{fig_1b}.  A word of caution is that if the parameters $\theta$ are not equal to the correct ones, the algorithm might not converge, and indeed sometimes does not.  But even in that case, the messages after $t_{\rm max}$ iterations can provide useful information. 

To conclude, we summarize some properties of the algorithm. At the correct parameters, the BP algorithm works in time linear in the size of the network, and is typically much faster than an equivalent Gibbs sampling algorithm.  Its superiority with respect to other community detection algorithms lies in the fact that, in addition to finding the group assignment that maximizes the overlap with the original assignment, it provides the marginal probabilities that each node belongs to each group, along with natural measures of the significance of the inferred assignment. 

Of course, here we assumed that we already knew the parameters $\theta$ with which the network was generated.  The next step is to use the BP algorithm as a subroutine to learn them, and to discuss under what circumstances this learning task is possible.

\subsection{Belief propagation algorithm to learn the parameters}

The Nishimori conditions  (\ref{na_Nish}--\ref{aa_Nish}) that we use for iterative learning can be written in terms of the BP messages as (for undirected graphs)
\bea
        n_a &=& \frac{1}{N}\sum_i \psi^i_a\, , \label{Nish_na_BP}\\
        c_{ab} &=& \frac{1}{N}  \frac{1}{n_b n_a}  \sum_{(i,j) \in E} \frac{c_{ab} ( \psi_a^{i\to j} \psi_b^{j\to i}+ \psi_b^{i\to j} \psi_a^{j\to i} )}{Z^{ij}}\, , \label{Nish_aa_BP}
\eea
where $Z^{ij}$ is defined in~(\ref{Z_ij}).  Therefore, BP can also be used to learn the optimal parameters as follows.

\begin{codebox}
\Procname{$\proc{BP-learning}(q,n^{\rm init}_a,c^{\rm init}_{ab},A_{ij},{\rm crit_{infer}},{\rm crit_{learn}})$}
\li $n_a \gets n^{\rm init}_a$, $c_{ab} \gets c^{\rm init}_{ab}$;
\li ${\rm conv} \gets {\rm crit_{learn}}+10$;
\li \While ${\rm conv}>{\rm crit_{learn}}$:
\li     \Do 
    $\proc{BP-inference}(q,n_a,c_{ab},A_{ij},{\rm crit_{infer}})$
\li Update $n_a$ and $c_{ab}$ according to (\ref{Nish_na_BP}--\ref{Nish_aa_BP}); \label{line_5}
\li ${\rm conv} \gets \sum_a |n_a^{\rm new}-n_a^{old} | +\sum_{ab} |c_{ab}^{\rm new}-c_{ab}^{old} | $
\End
\li \Return $n_a,c_{ab}$; 
\li \Return free energy; 
\end{codebox}

This is an expectation-maximization (EM) learning algorithm~\cite{DempsterLaird77}, where we use BP for the expectation step.  The update on line~\ref{line_5} can be also done using Gibbs sampling, but BP is much faster at computing the marginals.  The number of iterations needed for the EM algorithm to converge is constant in the size of the system.  However, it generally only converges to the correct $\theta$ from some finite fraction of the possible initial parameters $\theta^{\rm init}$.  
In practice, several initial values $\theta^{\rm init}$ need to be tried, and the fixed point with the smallest final free energy is the correct one.  

When the learning process is possible, this algorithm learns the group sizes $n_a$ and the affinity matrix $c_{ab}$ exactly in the limit $N \to \infty$.  To learn the number of groups $q$, we run \proc{BP-learning} for different values of $q$ and find the smallest $q^*$ such that the free energy density does not decrease further for larger $q$.

\section{Phase transition in inference and learning}
\label{fact_anal}

In this section we will limit ourselves to a particularly algorithmically difficult case of the block model, where the graph is undirected and every group $a$ has the same average degree $c$:
\be
  \sum_{d=1}^q c_{ad}n_d = \sum_{d=1}^q c_{bd}n_d = c \, ,  \quad {\rm for\,  all\, } a,b \, . \label{special}
\ee
If this is not the case, we can achieve a positive overlap with the original group assignment simply by labeling nodes based on their degrees, 
as we will briefly discuss in Section~\ref{sec_gen}. We call a block model satisfying (\ref{special}) a \emph{factorized block model}, and explain the reason for this name in the next paragraph.  Note that this case includes both the planted partitioning and the planted (noisy) coloring problem discussed in Section~\ref{model_def}.

The first observation to make about the belief propagation equations (\ref{BP_iter}) in the factorized block model is that 
\be 
\psi_{t_i}^{i\to j}=n_{t_i} \label{fact}
\ee 
is always a fixed point, as can be verified by plugging (\ref{fact}) into (\ref{BP_iter}). In the literature, a fixed point where messages do not depend on the indexes $i,j$ is called a \emph{factorized fixed point}, hence our name for this case of the block model. The free energy density at this fixed point is
\be
     f_{\rm factorized} =  \frac{c}{2}\left(1-\log{c}\right)  \, .\label{free_BP}
\ee 
For the factorized fixed point we have $\psi_{t_i}^i=n_{t_i}$, in which case the overlap (\ref{over_best}) is $Q=0$.  This fixed point does not provide any information about the original assignment---it is no better than a random guess.  If this fixed point gives the correct marginal probabilities and the correct free energy, we have no hope of recovering the original group assignment.  For which values of $q$ and $c_{ab}$ is this the case?

\subsection{Phase transitions in community detection}
\label{sec_stab}

We will first study the result given by the cavity method in the thermodynamic limit in the case when the parameters $q, \{n_a\}, \{c_{ab}\}$ used to generate the network are known. 

\begin{figure}[!ht]
   \includegraphics[width=0.495\linewidth]{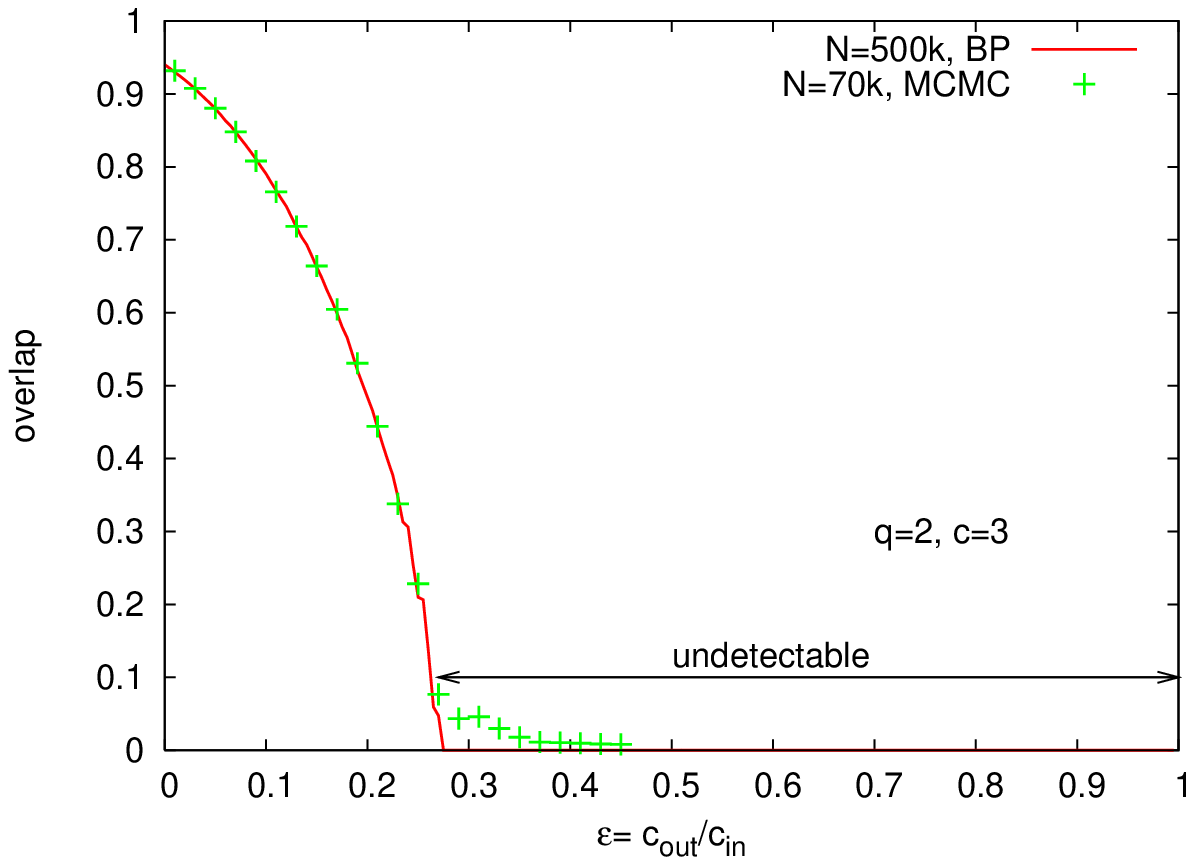}
   \includegraphics[width=0.495\linewidth]{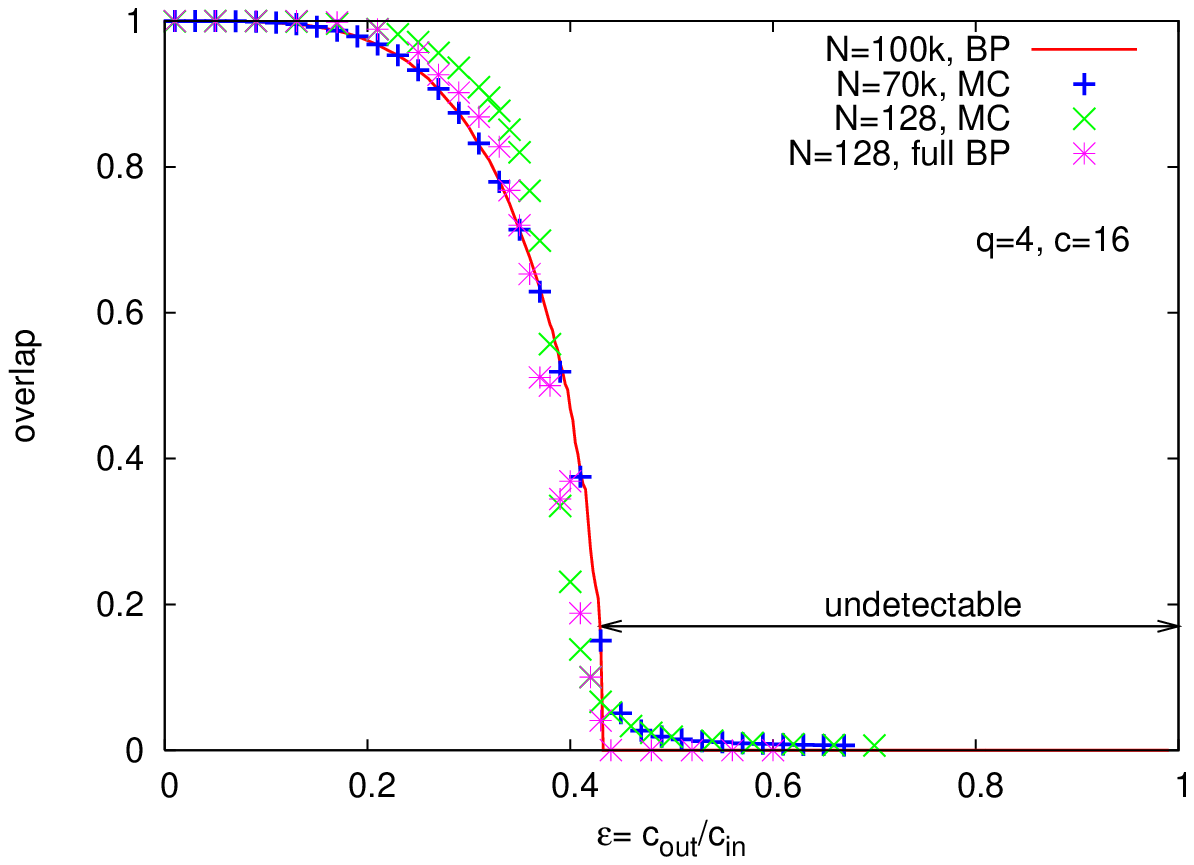}
   \caption{\label{fig_1} (color online): The overlap (\ref{overlap}) between the original assignment and its best estimate given the structure of the graph, computed by the marginalization (\ref{marginalization}). Graphs were generated using $N$ nodes, $q$ groups of the same size, average degree $c$, and different ratios $\epsilon=\cout/\cin$.  Thus $\epsilon=1$ gives an Erd\H{o}s-R\'enyi random graph, and $\epsilon=0$ gives completely separated groups.  Results from belief propagation (\ref{BP_iter}) for large graphs (red line) are compared to Gibbs sampling, i.e., Monte Carlo Markov chain (MCMC) simulations (data points).  The agreement is good, with differences in the low-overlap regime that we attribute to finite size fluctuations. On the right we also compare to results from the full BP (\ref{BP_iter_exact}) and MCMC for smaller graphs with $N=128$, averaged over $400$ samples. The finite size effects are not very strong in this case, and BP is reasonably close to the exact (MCMC) result even on small graphs that contain many short loops.  For $N\to \infty$ and $\epsilon> \epsilon_c= (c-\sqrt{c})/[c+\sqrt{c}(q-1)]$ it is impossible to find an assignment correlated with the original one based purely on the structure of the graph. For two groups and average degree $c=3$ this means that the density of connections must be $\epsilon_c^{-1}(q=2,c=3)=3.73$ greater within groups than between groups to obtain a positive overlap. For Newman and Girvan's benchmark networks with four groups (right), this ratio must exceed $2.33$.}
\end{figure}

\begin{figure}[!ht]
   \includegraphics[width=0.495\linewidth]{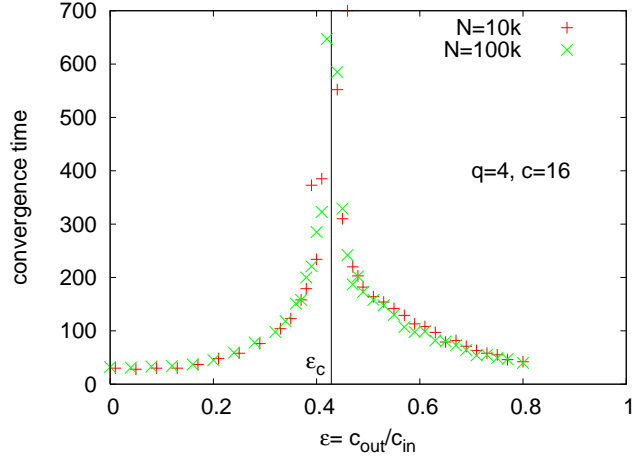}
   \caption{\label{fig_1b} (color online): The number of iterations needed for convergence of the BP algorithm for two different sizes. The convergence time diverges at the critical point $\epsilon_c$.  The equilibration time of Gibbs sampling (MCMC) has qualitatively the same behavior, but BP obtains the marginals much more quickly.}
\end{figure}

Fig.~\ref{fig_1} represents two examples where the overlap $Q$ is computed on a randomly generated graph with $q$ groups of the same size and an average degree $c$. We set $c_{aa}=\cin$ and $c_{ab}=\cout$ for all $a\neq b$ and vary the ratio $\epsilon=\cout/\cin$. The continuous line is the overlap resulting from the BP fixed point obtained by converging from a random initial condition (i.e., where for each $i,j$ the initial messages $\psi_{t_i}^{i\to j}$ are random normalized distributions on $t_i$). The convergence time is plotted in Fig.~\ref{fig_1b}. The points in Fig.~\ref{fig_1} are results obtained from Gibbs sampling, using the Metropolis rule and obeying detailed balance with respect to the Hamiltonian (\ref{Ham}), starting with a random initial group assignment $\{q_i\}$.  We see that $Q=0$ for $\cout/\cin > \epsilon_c$.  In other words, in this region both BP and MCMC converge to the factorized state, where the marginals contain no information about the original assignment.  For $\cout/\cin<\epsilon_c$, however, the overlap is positive and the factorized fixed point is not the one to which BP or MCMC converge. 

In particular the right-hand side of Fig.~\ref{fig_1} shows the case of $q=4$ groups with average degree $c=16$, corresponding to the benchmark of Newman and Girvan~\cite{NewmanGirvan04}.  We show the large $N$ results and also the overlap computed with MCMC for size $N=128$ which is the commonly used size for this benchmark.  Again, up to symmetry breaking, marginalization achieves the best possible overlap that can be inferred from the graph by any algorithm.  Therefore, when algorithms are tested for performance, their results should be compared to Fig.~\ref{fig_1} instead of to the common but wrong expectation that the four groups are detectable for any $\epsilon<1$.

Let us now investigate the stability of the factorized fixed point under random perturbations to the messages when we iterate the BP equations.  In the sparse case where $c_{ab}=O(1)$, graphs generated by the block model are locally treelike in the sense that almost all nodes have a neighborhood which is a tree up to distance $O(\log N)$, where the constant hidden in the $O$ depends on the matrix $c_{ab}$.  Equivalently, for almost all nodes $i$, the shortest loop that $i$ belongs to has length $O(\log N)$.  Consider such a tree with $d$ levels, in the limit $d\to \infty$. Assume that on the leaves the factorized fixed point is perturbed as 
\be
     \psi_{t}^{k} = n_t + \epsilon_{t}^{k} \, , 
\ee
and let us investigate the influence of this perturbation on the message on the root of the tree, which we denote $k_0$. There are, on average, $c^d$ leaves in the tree where $c$ is the average degree.  The influence of each leaf is independent, so let us first investigate the influence of the perturbation of a single leaf $k_d$, which is connected to $k_0$ by a path $k_d,k_{d-1},\dots,k_1,k_0$. We define a kind of transfer matrix 
\be
  T^{ab}_i 
  \equiv \frac{\partial \psi_{a}^{k_i}}{\partial \psi_{b}^{k_{i+1}}}\Big|_{\psi_t=n_t}
  = \left. \left[\frac{ \psi_{a}^{k_i} c_{ab}}{  \sum_{r} c_{ar} \psi_{r}^{k_{i+1}}  } 
   - \psi_{a}^{k_i} \sum_{s} \frac{ \psi_{s}^{k_i} c_{sb} }{ \sum_{r} c_{sr} \psi_{r}^{k_{i+1}}  } \right] \right|_{\psi_t=n_t} = n_a\left(\frac{c_{a b}}{c} - 1\right) \, .\label{matrix}
\ee 
where this expression was derived from~(\ref{BP_iter}) to leading order in $N$. The perturbation $\epsilon_{t_0}^{k_0}$ on the root due to the perturbation $\epsilon_{t_d}^{k_d}$ on the leaf $k_d$ can then be written as
\be
   \epsilon_{t_0}^{k_0} = \sum_{\{t_i\}_{i=1,\dots,d}} \left[ \prod_{i=0}^{d-1} T^{t_i,t_{i+1}}_{i} \right] \epsilon^{k_d}_{t_d} \label{iter_eps}
\ee
We observe in (\ref{matrix}) that the matrix $T^{ab}_i$ does not depend on the index $i$. Hence (\ref{iter_eps}) can be written as $\epsilon^{k_0} = T^d \epsilon^{k_d}$. When $d\to \infty$, $T^d$ will be dominated by $T$'s largest eigenvalue $\lambda$, so $\epsilon^{k_0} \approx \lambda^d \epsilon^{k_d}$. 

Now let us consider the influence from all $c^d$ of the leaves. The mean value of the perturbation on the leaves is zero, so the mean value of the influence on the root is zero. For the variance, however, we have
\be
      \left\langle  \left(\epsilon_{t_0}^{k_0}\right)^{\!2} \right\rangle 
      \approx \left\langle \left(\sum_{k=1}^{c^d}  \lambda^d \epsilon_{t}^{k}\right)^{\!\!2\,} \right\rangle  
      \approx c^d \lambda^{2d} \left\langle \left(\epsilon^k_{t}\right)^{\!2} \right\rangle \, .
\ee
This gives the following stability criterion, 
\be
        c \lambda^2 = 1 \, .   \label{stab}
\ee
For $c \lambda^2 < 1$ the perturbation on leaves vanishes as we move up the tree and the factorized fixed point is stable.  On the other hand, if $c \lambda^2 > 1$ the perturbation is amplified exponentially, the factorized fixed point is unstable, and the communities are easily detectable.

Consider the case with $q$ groups of equal size, where $c_{aa} = \cin$ for all $a$ and $c_{ab}=\cout$ for all $a\neq b$.  This includes the Newman-Girvan benchmarks, as well as planted (noisy) graph coloring and planted graph partitioning.  If there are $q$ groups, then $\cin + (q-1)\cout = qc$.  The transfer matrix $T^{ab}$ has only two distinct eigenvalues, $\lambda_1=0$ with eigenvector $(1,1,\ldots,1)$, and $\lambda_2=(\cin - \cout)/(qc)$ with eigenvectors of the form $(0,\dots,0,1,-1,0,\dots,0)$ and degeneracy $q-1$.   The factorized fixed point is then unstable, and communities are easily detectable, if 
\be 
|\cin - \cout|> q\sqrt{c}\, .  \label{KS}
\ee 

The stability condition (\ref{stab}) is known in the literature on spin glasses as the de Almeida-Thouless local stability condition~\cite{AlmeidaThouless78}, in information science as the Kesten-Stigum bound on reconstruction on trees~\cite{KestenStigum66,KestenStigum66b}, or the threshold for census reconstruction~\cite{MezardMontanari06}, or robust reconstruction threshold~\cite{JansonMossel04}.

We observed empirically that for random initial conditions both the belief propagation and the Monte Carlo Markov chain converge to the factorized fixed point when $c \lambda^2<1$. On the other hand when $c \lambda^2>1$ then BP and MCMC converge to a fixed point with a positive overlap, so that it is possible to find a group assignment that is correlated (often strongly) to the original assignment.  We thus conclude that if the parameters $q,\{n_a\},\{c_{ab}\}$ are known and if $c \lambda^2>1$, it is possible to reconstruct the original group assignment.

We can estimate the number of assignments that are just as good as the original one, i.e., that have the right group sizes and the right number of edges between each pair of groups, using Eq. (\ref{energy}) to express the entropy as 
\be
      s = \log{q} + \frac{c}{2}\log{c} - \frac{1}{2q} [(q-1) \cout \log{\cout} + \cin \log{\cin}] \, . \label{entropy}
\ee
We can think of this entropy as a measure of our uncertainty about the group assignment.

Next, we discuss a situation when the true marginal probabilities $\nu_i(t_i)$ are in the thermodynamic limit equal to $n_{t_i}$, and the free energy is given by~(\ref{free_BP}).  From the first fact it follows that the graph contains zero (or infinitesimally small) information about the original assignment; i.e., the overlap of the marginalized assignment with the original one is zero. To understand how is this possible even if $\epsilon\neq 1$, note that from the expressions for the free energy, it follows that the network generated with the block model is thermodynamically \emph{indistinguishable} from an Erd\H{o}s-R\'enyi random graph of the same average degree.  For the coloring problem where $c_{aa}=0$ and $(q-1)c_{ab}=qc$ for all $a\neq b$, this was proved in~\cite{AchlioptasCoja-Oghlan08} and discussed under the name ``quiet planting'' in~\cite{KrzakalaZdeborova09}. 

The main line of reasoning in the proof of~\cite{AchlioptasCoja-Oghlan08} goes as follows.  First note that the free energy (\ref{free_BP}) is equal to the annealed free energy
\be
         f_{\rm ann} = - \lim_{N\to \infty} \frac{1}{N} \log{[Z]_G } \, , 
\ee
where $[\cdot]_G$ denotes the average over the graphs. Then consider the following thought experiment.  First fix the parameters $q, \{n_a\}, \{c_{ab}\}$; then list in columns all the group assignments with group sizes $N_a=n_a N$; and list in rows all graphs with $M=cN/2$ edges.  Mark each (graph, assignment) pair $(G,\{q_i\})$ with the property that $G$ with assignment $\{q_i\}$ has the correct number of edges, $c_{ab} n_a n_b N$, between each pair of groups.  

Now consider two ways to choose these pairs.  The block model corresponds to choosing a random assignment (column), and then choosing a random graph (row) consistent with it.  In contrast, we can start by choosing a random graph (row), and then choose an assignment (column) consistent with the block model.  If there are exactly the same number of marked pairs in every row and column, these two procedures are the same.  In that case, it would be impossible to distinguish a graph generated by the block model from an Erd\H{o}s-R\'enyi graph.  

It is certainly true that for every group assignment (with fixed group sizes) there are the same number of compatible graphs.  But different graphs have different numbers of compatible assignments.  Thus the number of marked pairs is different in different rows.  The number of marked pairs in each row is essentially the partition function $Z$.  Now, if all but an exponentially small fraction of graphs have the same typical properties as a random graph (this is a large deviation principle), it follows that also the block model generates typical random graphs as long as the quenched free energy equals the annealed one, i.e., when $\lim_{N\to \infty}  \log{[ Z]_G }/N = \lim_{N\to \infty} [\log{ Z}]_G/N$. 

In the example presented in Fig.~\ref{fig_1}, both BP and MCMC confirm that the free energy (\ref{free_BP}) is the correct free energy and that the marginals are uniform, $\nu_i(a)=n_a$, for $\epsilon>\epsilon_c$. The only possibility known in statistical physics when MCMC does not give the correct free energy is ergodicity breaking on the time-scale during which the MCMC was performed (i.e. insufficient equilibration time). Indeed, the original group assignment could belong to a part of the phase space that dominates the Boltzmann distribution, but that is invisible to dynamics on time-scales linear in the size of the system. Such a situation indeed happens in systems that undergo the ideal glass transition. There is a simple trick to verify if such a glass transition appears or not in the block model: just run BP or MCMC using the original group assignment as the initial condition. 

When we do this for the examples shown in Fig.~\ref{fig_1}, there is absolutely no change in the result.  Thus the free energy and overlaps presented in Fig.~\ref{fig_1} are asymptotically exact and we can distinguish two phases:
\begin{itemize}
   \item{If $|\cin - \cout|< q\sqrt{c}$, the graph does not contain any significant information about the original group assignment, and community detection is  impossible.}
   \item{If $|\cin - \cout|> q\sqrt{c}$, the graph contains significant information about the original group assignment, and using BP or MCMC yields an assignment that is strongly correlated with the original one.  There is some intrinsic uncertainty about the group assignment due to the entropy, but if the graph was generated from the block model there is no better method for inference than the marginalization introduced by Eq.~(\ref{marginalization}). }
\end{itemize}
Fig.~\ref{fig_1} hence illustrates a phase transition in the detectability of communities.  Unless the ratio $\cout/\cin$ is far enough from $1$, the groups that truly existed when the network was generated are undetectable from the topology of the network.  Moreover, unless the condition (\ref{KS}) is satisfied the graph generated by the block model is indistinguishable from a random graph, in the sense that typical thermodynamic properties of the two ensembles are the same.

\begin{figure}[!ht]
 \includegraphics[width=0.495\linewidth]{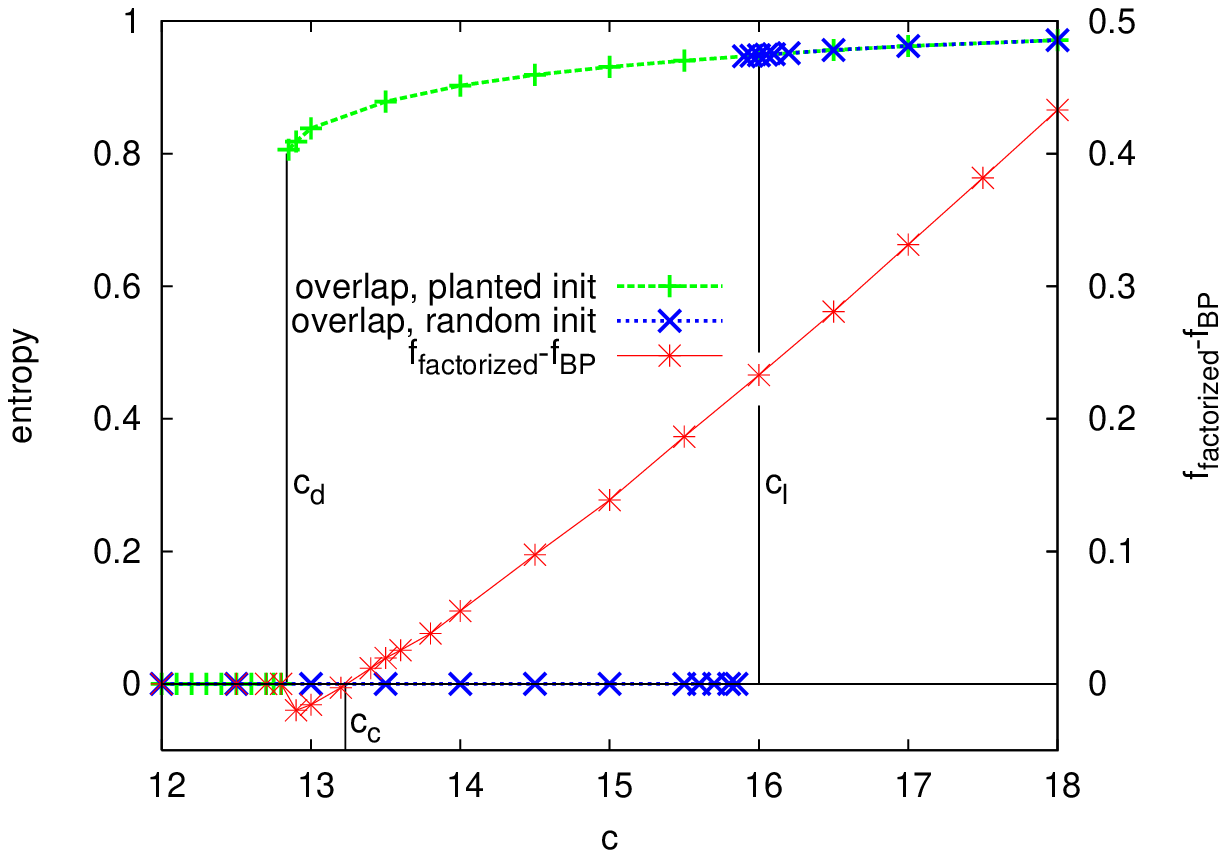}
 \includegraphics[width=0.495\linewidth]{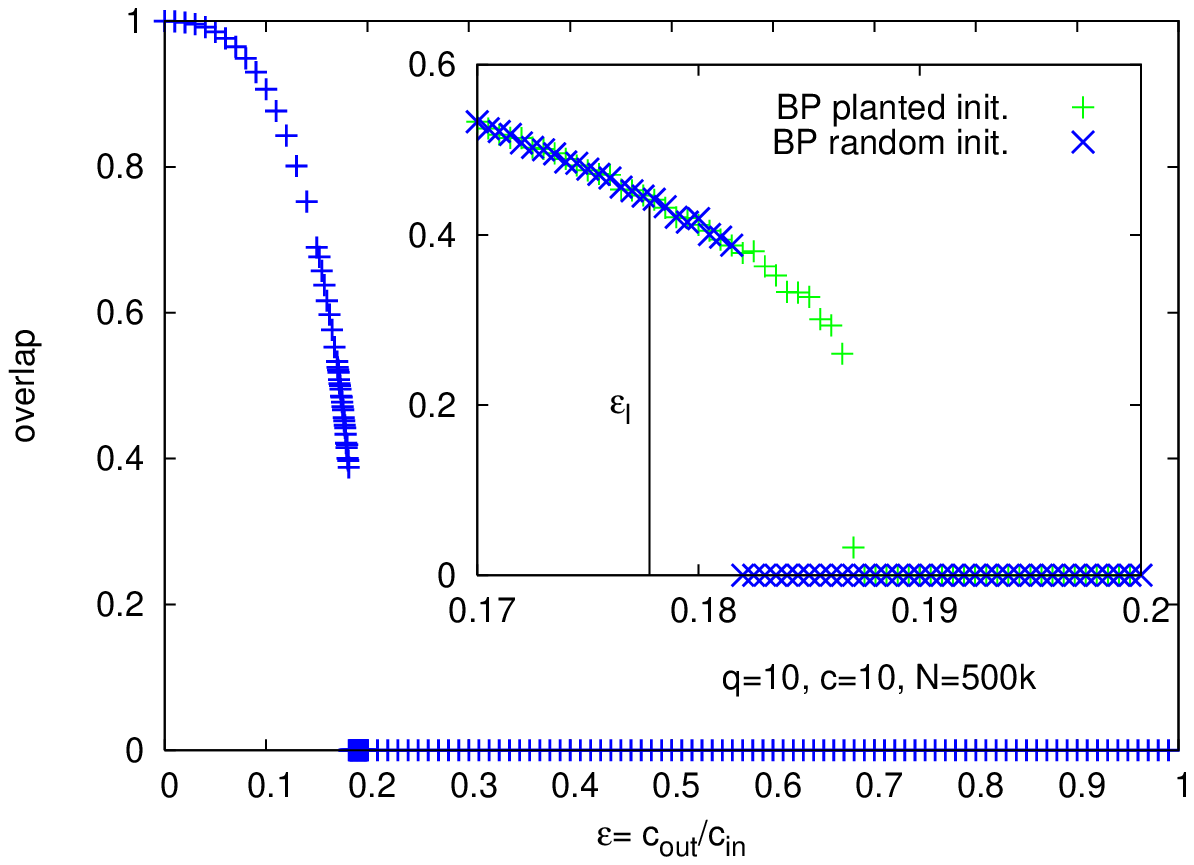}
  \caption{\label{fig_2} (color online): Left: graphs generated with $q=5$, $\cin=0$, and $N=10^5$. We compute the overlap (\ref{overlap}) and the free energy with BP for different values of the average degree $c$. The green crosses show the overlap of the BP fixed point resulting from using the original group assignment as the initial condition, and the blue crosses show the overlap resulting from random initial messages. The red stars show the difference between the factorized free energy (\ref{free_BP}) and the free energy resulting from the planted initialization. We observe three important points where the behavior changes qualitatively: $c_d=12.84$, $c_c=13.23$, and $c_\ell=16$.  We discuss the corresponding phase transitions in the text.
Right: the case $q=10$ and $c=10$.  We plot the overlap as a function of $\epsilon$; it drops down abruptly from about $Q=0.35$. The inset zooms in on  the critical region. We mark the stability transition $\epsilon_\ell$, and data points for $N=5\cdot 10^5$ for both the random and planted initialization of BP. In this case the data are not so clear. The overlap from random initialization becomes positive a little before the asymptotic transition. We think this is due to strong finite size effects. From our data for the free energy it also seems that the transitions $\epsilon_c$ and $\epsilon_d$ are very close to each other (or maybe even equal, even though this would be surprising). These subtle effects are, however, relevant only in a very narrow region of $\epsilon$ and are, in our opinion, not likely to appear for real-world networks.}
\end{figure}

The situation illustrated in Fig.~\ref{fig_1} is, however, not the most general one. Fig.~{\ref{fig_2}} illustrates the case of planted coloring with $q=5$, $\cin=0$, and $\cout=qc/(q-1)$. In this case the condition for stability (\ref{KS}) leads to a threshold value $c_\ell=(q-1)^2$. We plot again the overlap obtained with BP, using two different initializations: the random one, and the planted one corresponding to the original assignment.  In the latter case, the initial messages are
\be
      \psi_{q_i}^{i\to j} = \delta_{q_i t_i} \, , 
\ee
where $t_i$ is the original assignment. We also plot the corresponding BP free energies.  As the average degree $c$ increases, we see four different phases in Fig.~\ref{fig_2}:
\begin{itemize}
   \item[I.] For $c < c_d$, both initializations converge to the factorized fixed point, so the graph does not contain any significant information about the original group assignment.  The ensemble of assignments that have the proper number of edges between each pair of groups is thermodynamically indistinguishable from the uniform ensemble. The original assignment is one of these configurations, and there is no possible way to tell which one it is without  additional knowledge. 
   \item[II.] For $c_d < c < c_c$, the planted initialization converges to a fixed point with positive overlap, and its free energy is larger than the annealed free energy. In this phase there are exponentially many basins of attraction (states) in the space of assignments that have the proper number of edges between each pair of groups. These basins of attraction have zero overlap with each other, so none of them yield any information about any of the others, and there is no way to tell which one of them contains the original assignment.  The annealed free energy is still the correct total free energy, the graphs generated by the block model are thermodynamically indistinguishable from Erd\H{o}s-R\'enyi random graphs, and there is no way to find a group assignment correlated with the original one.  
   \item[III.] For $c_c < c < c_\ell$, the planted initialization converges to a fixed point with positive overlap, and its free energy  is smaller than the annealed free energy.  There might still be exponentially many basins of attraction in the state space with the proper number of edges between groups, but the one corresponding to the original assignment is the one with the largest entropy and the lowest free energy.  Therefore, if we can perform an exhaustive search of the state space, we can infer the original group assignment.  However, this would take exponential time, and initializing BP randomly almost always leads to the factorized fixed point. In this phase, inference is possible, but exponentially hard; the state containing the original assignment is, in a sense, hidden below a glass transition.  Based on the physics of glassy systems, we predict that no polynomial-time algorithm can achieve a positive overlap with the original group assignment. 
   \item[IV.] For $c > c_\ell$, both initializations converge to a fixed point with positive overlap, strongly correlated with the original assignment.  Thus inference is both possible and easy, and BP achieves it in linear time.  Indeed, in this easy phase, many efficient algorithms will be able to find a group assignment strongly correlated with the original one.  
\end{itemize}

We also investigated the case $q=5$, $\cin=0$, illustrated in Fig.~\ref{fig_2}, with Gibbs sampling, i.e., the Markov chain Monte Carlo algorithm.  For the planted initialization, its performance is generally similar to BP.  For the random initialization, MCMC agrees with BP only in phases (I) and (IV).  It follows from results on glassy systems~\cite{ZdeborovaKrzakala10} that in phases (II) and (III), the equilibration time of MCMC is exponentially large as a function of $N$, and that its performance in linear time, i.e., $CN$ for any constant $C$, does not yield any information about the original group assignment.

The boundaries between different phases correspond to well-known phase transitions in the statistical physics of spin glasses.  Specifically, $c_d$ is the dynamical transition or reconstruction threshold, see e.g.~\cite{MezardMontanari06,ZdeborovaKrzakala07}. The detectability threshold $c_c$ corresponds to the condensation transition or the Kauzmann temperature.  Finally, $c_\ell$ is the easy/hard transition in planted models introduced in~\cite{KrzakalaZdeborova09}. There is also a close relation between our approach and optimal finite temperature decoding~\cite{Nishimori93,Sourlas94,Iba99,NishimoriBook01} and the statistical mechanics approach to image processing~\cite{Tanaka02}. 

We saw in our experiments, consistent with the reconstruction thresholds $c_d$ and the Kesten-Stigum bound $c_\ell$ in~\cite{MezardMontanari06}, that for assortative communities where $\cin>\cout$, phases (II) and (III) are extremely narrow or nonexistent.  For $q \le 4$, these phases do not exist, and the overlap grows continuously from zero in phase (IV), giving a continuous phase transition as illustrated in Fig.~\ref{fig_1}.  For $q \ge 5$,  phases (II) and (III) occur in an extremely narrow region, as shown on the right in Fig.~\ref{fig_2}. The overlap jumps discontinuously from zero to a relatively large value, giving a discontinuous phase transition. In the disassortative (antiferromagnetic) case where $\cin > \cout$, phases (II) and (III) are more important.  For instance, when $\cin=0$ and the number of groups is large, the thresholds scale as $c_d \approx q \log{q}$, $c_c\approx 2  q \log{q}$ and $c_\ell=(q-1)^2$.  However, in phase (III) the problem of inferring the original assignment is as hard as finding a solution to a random Boolean satisfiability problem close to the satisfiability threshold, or coloring a random graph close to the $q$-colorability threshold.  These are NP-complete problems, and are believed to be exponentially hard in this region.

In phase (IV), i.e., when (\ref{KS}) is satisfied, inference is easy in linear time with both BP and MCMC, and likely with many other community detection algorithms in the literature (at least in the assortative case, which has received the most attention).  But, at least for networks generated by the block model, there is very little space for algorithmic improvement: either the inference problem is easy (in linear time), or exponentially hard, or impossible.  Our results suggest that there is no middle ground where inference is possible but requires time, say, $O(N^c)$ for some $c > 1$.

As far as we know, the phase transitions presented here are previously unknown in the literature on community detection.  However, the phase transition in detectability was predicted by the authors of~\cite{ReichardtLeone08}.  They used an approximate (replica symmetric) calculation of the ground state energy of the Hamiltonian (\ref{Ham}) on networks created by the block model, and noticed that it differs from its value on random graphs only when $\cout/\cin$ is sufficiently small.  
Based on this, they predicted an undetectable region if the probability that a random edge connects two nodes of the same community is smaller than a critical value $p_c$.  Our exact calculation (\ref{KS}) leads to $p_c = [c+(q-1)\sqrt{c}]/(qc)$ when $\cin>\cout$, showing that~\cite{ReichardtLeone08} overestimated the size of the undetectable region; this also explains the discrepancy with their numerical data.  The problem with their calculation is that the ground state energy of (\ref{Ham}) cannot be computed correctly using the replica symmetric approximation, and that the ground state does not maximize the overlap with the original assignment in any case.  In contrast, we focus on the free energy, and our calculations are exact in the thermodynamic limit.  In addition, \cite{ReichardtLeone08} only treated the cases in Fig.~\ref{fig_1}, and did not encounter the case when the detectability transition is discontinuous.  

Another related work that numerically investigated the behavior of a Potts-like Hamiltonian (\ref{Ham}) with general parameters on networks generated from the block model is~\cite{HuRonhovde10}. The principal difference between~\cite{HuRonhovde10} and our work is that they do not focus on the parameters of the Hamiltonian with which the network was generated, or discuss optimal inference of the group assignment with those parameters.

\subsection{Phase transitions in parameter learning}
\label{fact_learning}

In this section we will continue to focus on the special case of the parameters defined by (\ref{special}), where all groups have the same average degree.  We will, however, no longer assume that the correct values of the parameters $q$, $\{n_a\}$ and $\{c_{ab}\}$ are known; now our goal is to learn them.

\begin{figure}[!ht]
\includegraphics[width=0.495\linewidth]{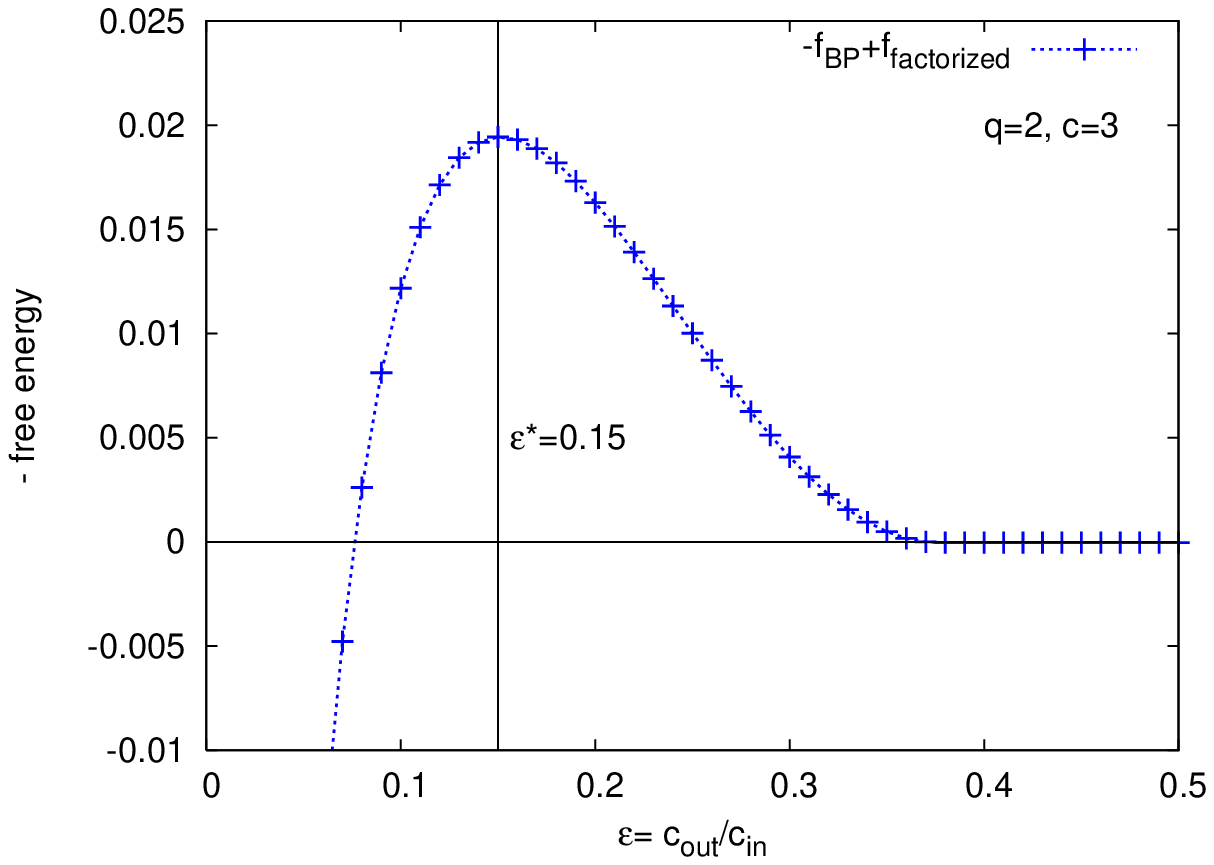} \includegraphics[width=0.495\linewidth]{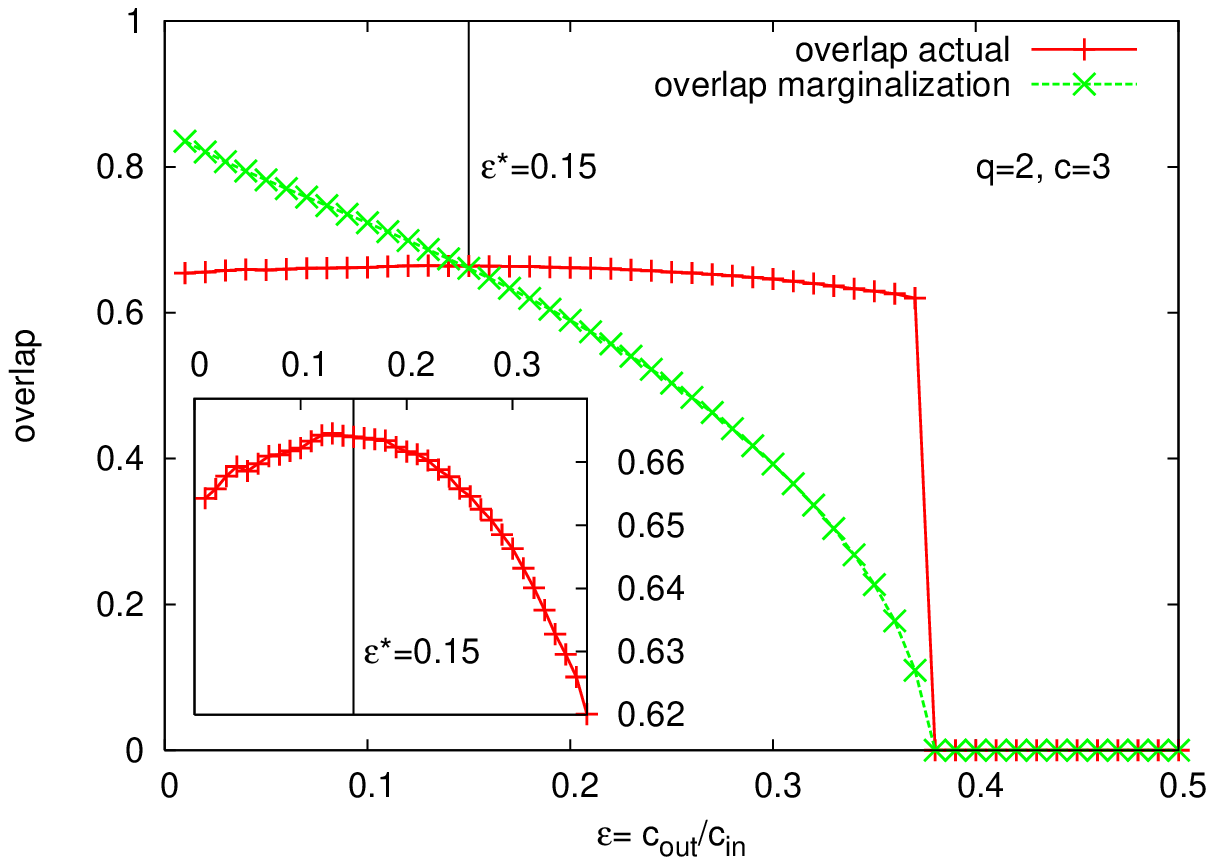}
  \caption{\label{fig_3} (color online): Learning for graphs of $N=10^5$ nodes with $q=2$ groups, average degree $c=3$, and $\epsilon^*=\cout/\cin=0.15$. Left: the BP free energy as a function of $\epsilon$.  Specifically, we plot the factorized free energy (which is independent of $\epsilon$) minus the BP free energy. As we expect, the maximum is achieved at $\epsilon=\epsilon^*$.  Our learning procedure looks for this maximum via a kind of expectation-maximization (EM) algorithm.  
Note that for $\epsilon> \epsilon_s = 0.36$ the BP free energy is equal to the factorized one, so we need to initialize the learning process somewhere in the region $\epsilon<\epsilon_s$. Right: the overlap (\ref{overlap}) between the original group assignment and the best estimate using BP marginalization, compared to the estimated overlap (\ref{over_best}). They are equal only at the correct parameters, $\epsilon=\epsilon^*$. In the inset we see that the actual overlap is maximized at $\epsilon=\epsilon^*$, illustrating that to infer the group assignment optimally one needs to have the correct parameters.}
\end{figure}

In Fig.~\ref{fig_3}, we generate graphs with $q=2$ groups of the same size, $N=10^5$, with average degree $c$, and $c_{11}=c_{22}=\cin$ and $c_{12}=\cout$ where $\epsilon^*=\cout/\cin=0.15$. We then compute the free energy as a function of $\epsilon$. 
The left-hand side of Fig.~\ref{fig_3} shows the factorized free energy (\ref{free_BP}) minus the free energy obtained from the BP fixed point as a function of $\epsilon$.  As expected, this curve is maximized at the correct value $\epsilon=\epsilon^*$.  The learning procedure searches for this maximum.  

Note that the initial values of the parameters are important.  For $\epsilon>\epsilon_s=0.36$, the free energy equals the factorized one, and BP converges to the factorized fixed point. Hence if we start the learning process with $\epsilon>\epsilon_s$, the local slope of the free energy will not push us in the correct direction.  

The right-hand side of Fig.~\ref{fig_3} shows (red crosses) the corresponding values of the overlap between the marginalized group assignment and the original one.  We see that this overlap is maximized at the correct value $\epsilon=\epsilon^*$.  We also plot the estimated overlap (Eq.~(\ref{over_best}), green crosses), which is based only on the marginals with no knowledge of the original group assignment.  At $\epsilon=\epsilon^*$, it indeed equals the true overlap with the original assignment.

\begin{figure}[!ht] \includegraphics[width=0.495\linewidth]{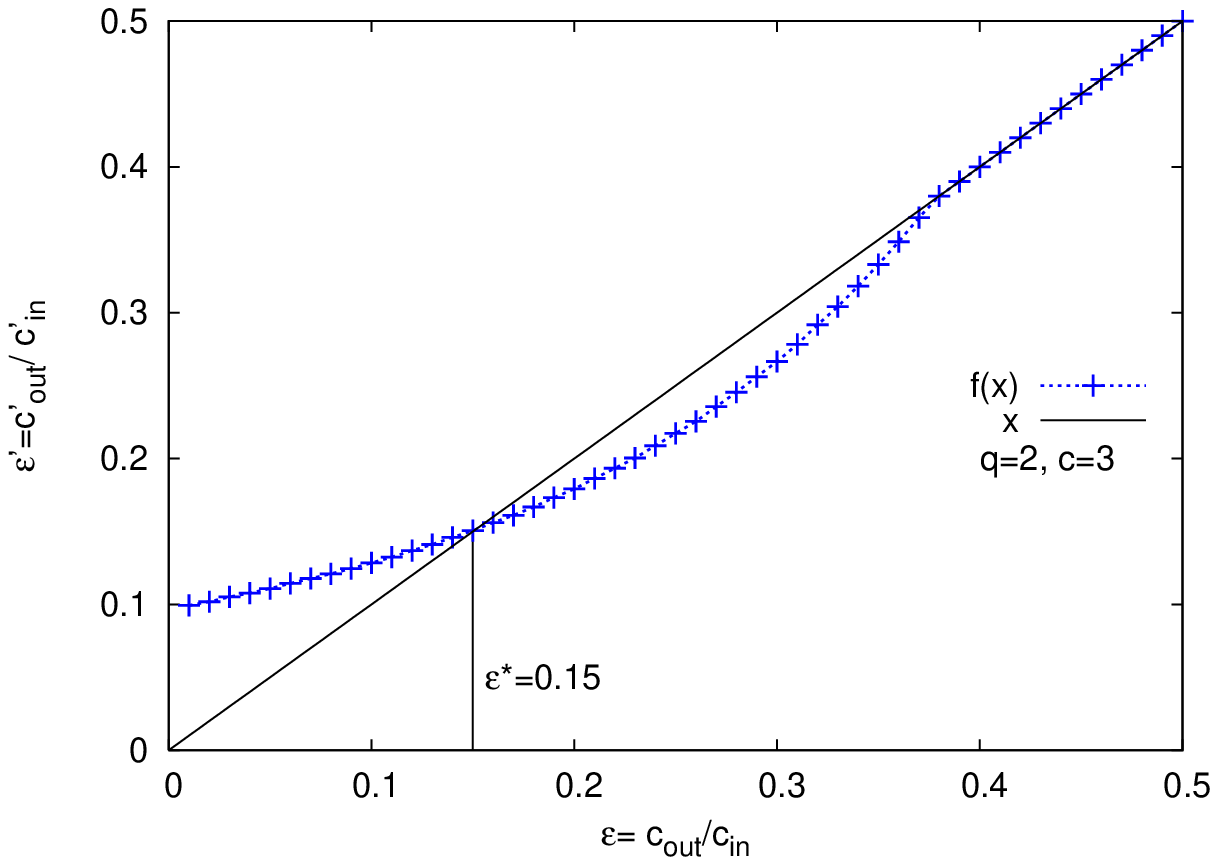}
\includegraphics[width=0.495\linewidth]{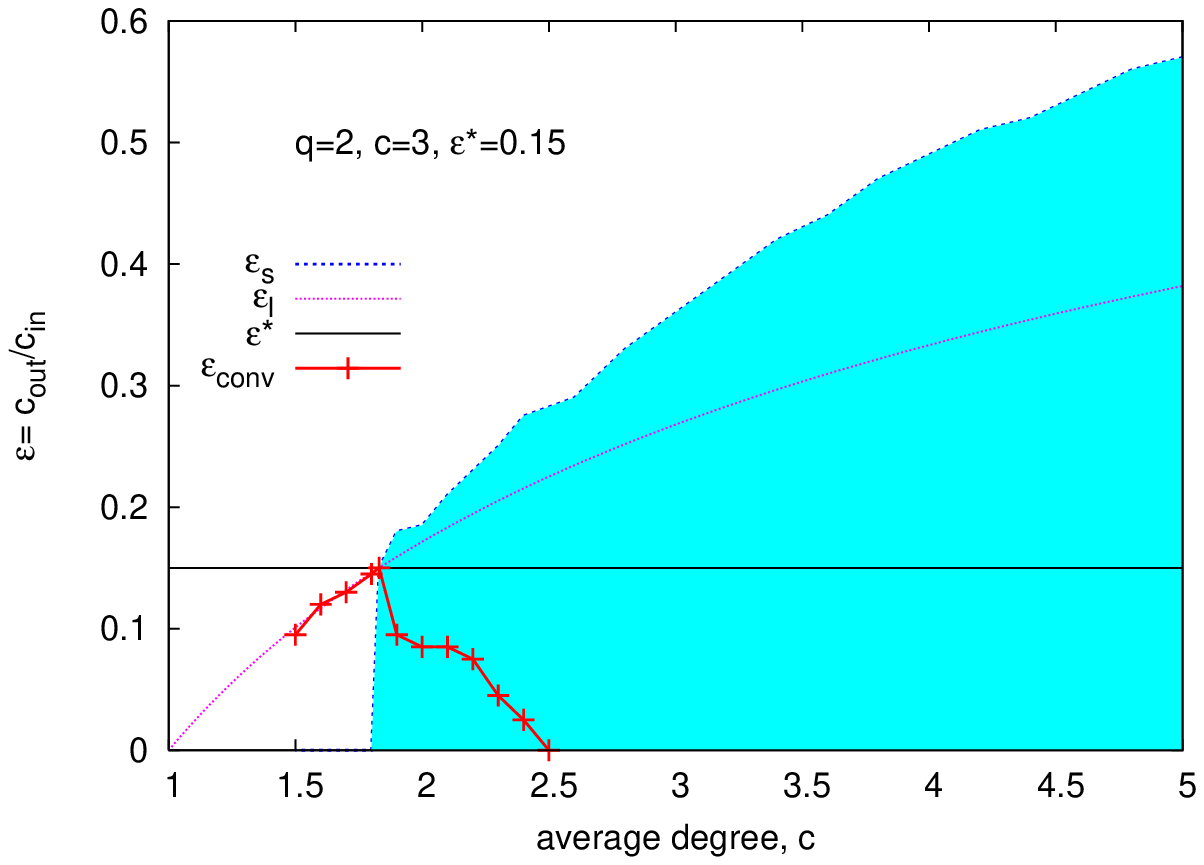}
  \caption{\label{fig_4}  (color online): Left: learning on graphs generated with $q=2$, $c=3$, and $\epsilon^*=\cout/\cin=0.15$.  For each $\epsilon$ we compute the averages in (\ref{ab_Nish}--\ref{aa_Nish}) from the BP fixed point, update $\cout$ and $\cin$ accordingly, and plot the new ratio $\epsilon'=\cout'/\cin'$ as a function of $\epsilon$. This process converges to $\epsilon^*$ if we initialize it at $\epsilon < \epsilon_s$: in contrast, every $\epsilon > \epsilon_s$ is an (incorrect) fixed point.  Right: the shaded region illustrates the initial values $(\epsilon,c)$ of the parameters from which the learning process converges to $\epsilon^*$.  Learning is possible for $c > c_\ell$, where $c_\ell$ is given by (\ref{KS}).  Graph generated with  $q=2$, $c=3$, $\epsilon^*=0.15$, and different values of average degree $c$. BP is run with $\epsilon\neq \epsilon^*$, for $\epsilon<\epsilon_{\rm conv}$ the BP does not converge.  
The magenta line corresponds to the largest $\epsilon^*(c)$, given by~\eqref{KS}, for which communities are detectable at a given average degree $c$. 
}
\end{figure}

Fig.~\ref{fig_4} uses graphs generated in the same way as in Fig.~\ref{fig_3}. For each $\epsilon$ we compute the averages (\ref{ab_Nish}--\ref{aa_Nish}) from the BP fixed point.  In terms of the BP messages, the most likely values of the parameters are then, as in~\eqref{Nish_aa_BP},
\bea
        \cout' &=& \frac{q^2}{N} \sum_{(i,j) \in E} \frac{\cout ( \psi_1^{i\to j} \psi_2^{j\to i}+ \psi_2^{i\to j} \psi_1^{j\to i} )}{  Z^{ij} } \label{iter_out}\\
        \cin' &=& \frac{2q^2}{N}  \sum_{(i,j) \in E} \frac{ \cin\psi_1^{i\to j} \psi_1^{j\to i}}{Z^{ij}} \, . \label{iter_in}
\eea
The learning process iteratively updates $\cin$ and $\cout$ (more generally, the affinity matrix $c_{ab}$) and looks for a fixed point where $\epsilon' = \epsilon$.  As we said above, this is essentially an expectation-maximization (EM) algorithm, where we use BP to approximate the expectation step.

In Fig.~\ref{fig_4} we plot $\epsilon'=\cout'/\cin'$ as a function of $\epsilon$. We see that $\epsilon^*$ is the only fixed point in its vicinity.  However, every $\epsilon>\epsilon_s$ is also a fixed point due to the factorized BP fixed point, again showing that we need to initialize the learning process at some $\epsilon < \epsilon_s$.

On the right-hand side of Fig.~\ref{fig_4} we depict the region in the $(\epsilon,c)$ plane in which the learning process converges to $\epsilon^*$ if we start it at $\epsilon$.  We see that learning is possible for $c>c_\ell=1.83$, where $c_\ell$ was obtained from (\ref{KS}) by considering $\epsilon^*=0.15$. But even in this region $c>c_\ell$ one should not start with too large a value of $\epsilon$.  It is better to start with $\epsilon=0$ (i.e., completely separated groups) rather than with $\epsilon=1$ (an undifferentiated random graph). The same is true in the antiferromagnetic case, i.e., if $\cout>\cin$, if we define $\epsilon=\cin/\cout$. 

In general we conclude that the phase transitions for inference (i.e. when parameters are known) are present also in learning. Whenever inference is possible the asymptotically correct parameters can be learned with a proper initialization. The set of good initial values of the parameters always takes a finite fraction of the all possible initial values (hence finding good initialization takes a finite number of steps).  

\begin{figure}[!ht]
  \includegraphics[width=0.495\linewidth]{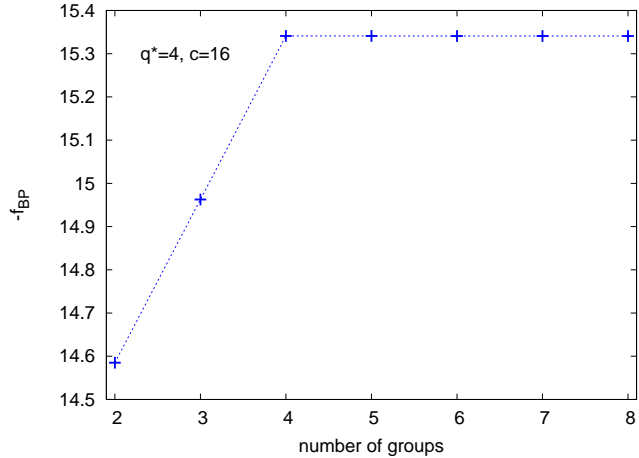}
  \caption{\label{fig_5} (color online): The (negative) free energy for a graph generated with $q^*=4$ groups, average degree $c=16$, $\epsilon=0.2$, and $N=10^4$.  We run BP for various values of $q$ and plot the (negative) free energy. The correct number of groups is the $q$ at which the free energy saturates.}
\end{figure}

If the group sizes are unknown, we can learn them in a similar manner, updating them using~\eqref{Nish_na_BP}. On the other hand, learning the number of groups requires a different approach.  Fig.~\ref{fig_5} shows the dependence of the free energy on $q$, for an example where the correct number of groups is $q^*=4$.  If $q > q^*$, there are multiple assignments where $q-q^*$ groups are empty, so the free energy is not maximized at $q^*$.  Instead, the free energy grows as long as $q<q^*$, and then stays constant for $q \ge q^*$. To learn the correct number of groups we thus need to run the algorithm for several values of $q$ and select the $q$ at which the free energy stops growing. 

It is instructive to discuss the parameter values that achieve the maximum free energy when $q>q^*$. These are for instance group sizes where $q-q^*$ groups are empty. But there is, in general, a continuum of other fixed points of the learning process with the same free energy; for instance, where one group is divided into two in an arbitrary way.  The learning process converges to one of these fixed points, and we have not found a way to determine $q^*$ more directly than running BP with a lower value of $q$ and comparing the free energies.  We stress that this method of learning the number of groups is asymptotically exact for networks generated by the block model.  However, for real networks the free energy of the block model does not generally saturate at any finite $q$, as we will discuss in the next section.

\subsection{When groups have unequal average degree}
\label{sec_gen}

In the previous section we studied the asymptotic behavior of the stochastic block model in the case (\ref{special}) where every group has the same average degree.  In that case, the degree sequence does not contain any information about which node should be in which group, and phases exist in parameter space where the original group assignment cannot be inferred from the structure of the network.

If (\ref{special}) is not satisfied, each group $a$ has average degree $c_a$ depending on $a$.  In that case, the undetectable phase does not exist, since classifying nodes according to their degree yields a nonzero overlap with the original group assignment.  Our procedure for optimal inference and parameter learning described in Sections~\ref{bayes_inference} and~\ref{bayes_learning} still holds here, as do the BP algorithm and its asymptotic analysis.  Section~\ref{bp_dep} can be used to infer the original assignment and the model parameters, and these algorithms are asymptotically exact.  In the generic case where the group average degrees are all distinct, they can be learned exactly in the thermodynamic limit, even if the differences between them are small.

The phase transitions in the inference problem described in Section~\ref{sec_stab}, however, still exist and can again be investigated using the cavity method, although the condition (\ref{KS}) for easy inference does not have a simple analytic form anymore.  In the case where the detectability phase transition is discontinuous, we can again analyze the phase diagram by considering BP with the planted and random initializations. As we travel along some curve through the space of parameters $\{n_a\}, \{c_{ab}\}$, 
the transition $c_d$ corresponds to the point where the two initializations start to converge to two different fixed points, $c_c$ (the detectability transition) to the point at which their free energies become equal, and finally $c_\ell$ (the hard/easy transition) to the point where they both converge to the same fixed point, which is strongly correlated with the original assignment.  In the case where the phase transition is continuous, there is just one transition, where the BP convergence time diverges as in Fig.~\ref{fig_1b}.

\section{Performance on real-world networks}
\label{sec_real}

We tested our inference and learning algorithms on several real-world networks.  We present our findings on two particular examples: Zachary's karate club~\cite{Zachary77} and a network of books on politics.  
The purpose of this discussion is not to argue that our algorithms outperform other known methods; both these networks are small, with easily-identifiable communities.  Rather, our point is that our algorithms provide a 
quantitative comparison between these real-world networks and those generated by the stochastic block model.  
More generally, our techniques allow us to quantitatively study the extent to which a network is well-modeled by a given generative model, a study that we feel more work should be devoted to in the future.

First let us make a remark about our algorithm's performance on synthetic benchmarks that are generated by the stochastic block model.  The results on the right-hand side of Fig.~\ref{fig_1} correspond to the four-group networks of Newman and Girvan~\cite{NewmanGirvan04}, that have been used as benchmarks for many algorithms in the literature.  Up to symmetry breaking, the overlap with the original group assignment shown in Fig.~\ref{fig_1} is the best that can be achieved by any inference algorithm, and both MCMC (Gibbs sampling) and the BP algorithm achieve this optimum in linear time.  Thus the right way to measure the performance of a community detection algorithm on these networks is to compare their results to Fig.~\ref{fig_1}.  This holds also for more general synthetic benchmarks generated by the block model, like those in~\cite{LancichinettiFortunato08}.

\subsection{Zachary's karate club}

Zachary's karate club~\cite{Zachary77} is a popular example for community detection.  It consists of friendships between the members of a karate club which split into two factions, one centered around the club president and the other around the instructor.  It has $34$ nodes and $78$ edges.  We ran the BP learning algorithm on this network with $q=2$ groups.  Depending on the initial parameters $\{n_a\}, \{c_{ab}\}$, it converges to one of two attractive fixed points in parameter space:
\bea
  n^{(i)} &=&
  \left(
   \begin{array}{c}
     0.525 \\
     0.475
   \end{array} 
  \right) \; , \;\quad 
  c^{(i)} = 
   \left(
   \begin{array}{cc}
     8.96 & 1.29 \\
     1.29 & 7.87
   \end{array} 
  \right) \, ,
\nonumber \\
  n^{(ii)} &=&
  \left(
   \begin{array}{c}
     0.854 \\
     0.146
   \end{array} 
  \right) \; , \; \quad
  c^{(ii)} = 
   \left(
   \begin{array}{cc}
     16.97 & 12.7 \\
     12.7 & 1.615
   \end{array} 
  \right) \, .
  \label{bp-fixedpoints}
\eea
For comparison, we also performed learning using MCMC for the expectation step; this network is small enough, with such a small equilibration time, that MCMC is essentially exact.  We again found two attractive fixed points in parameter space, very close to those in~\eqref{bp-fixedpoints}:
\bea
  n_{\rm MC}^{(i)} =
  \left(
   \begin{array}{c}
     0.52 \\
     0.48
   \end{array} 
  \right) \; , \; \quad
  c_{\rm MC}^{(i)} = 
   \left(
   \begin{array}{cc}
     8.85 & 1.26 \\
     1.26 & 7.97
   \end{array} 
  \right) \, ,
\nonumber \\
  n_{\rm MC}^{(ii)} =
  \left(
   \begin{array}{c}
     0.85 \\
     0.15
   \end{array} 
  \right) \; , \;\quad
  c_{\rm MC}^{(ii)} = 
   \left(
   \begin{array}{cc}
     16.58 & 12.52 \\
     12.52 & 1.584
   \end{array} 
  \right) \, .
\eea

A first observation is that even though Zachary's karate club is both small and ``loopy,'' rather than being locally treelike, the BP algorithm converges to fixed points that are nearly the same as the (in this case exact) MCMC.  This is despite the fact that our analysis of the BP algorithm assumes that there are no small loops in the graph, and focuses on the thermodynamic limit $N \to \infty$.  This suggests that our BP learning algorithm is a useful and robust heuristic even for real-world networks that have many loops.

\begin{figure}[!ht]
   \includegraphics[angle=90,width=0.40\linewidth]{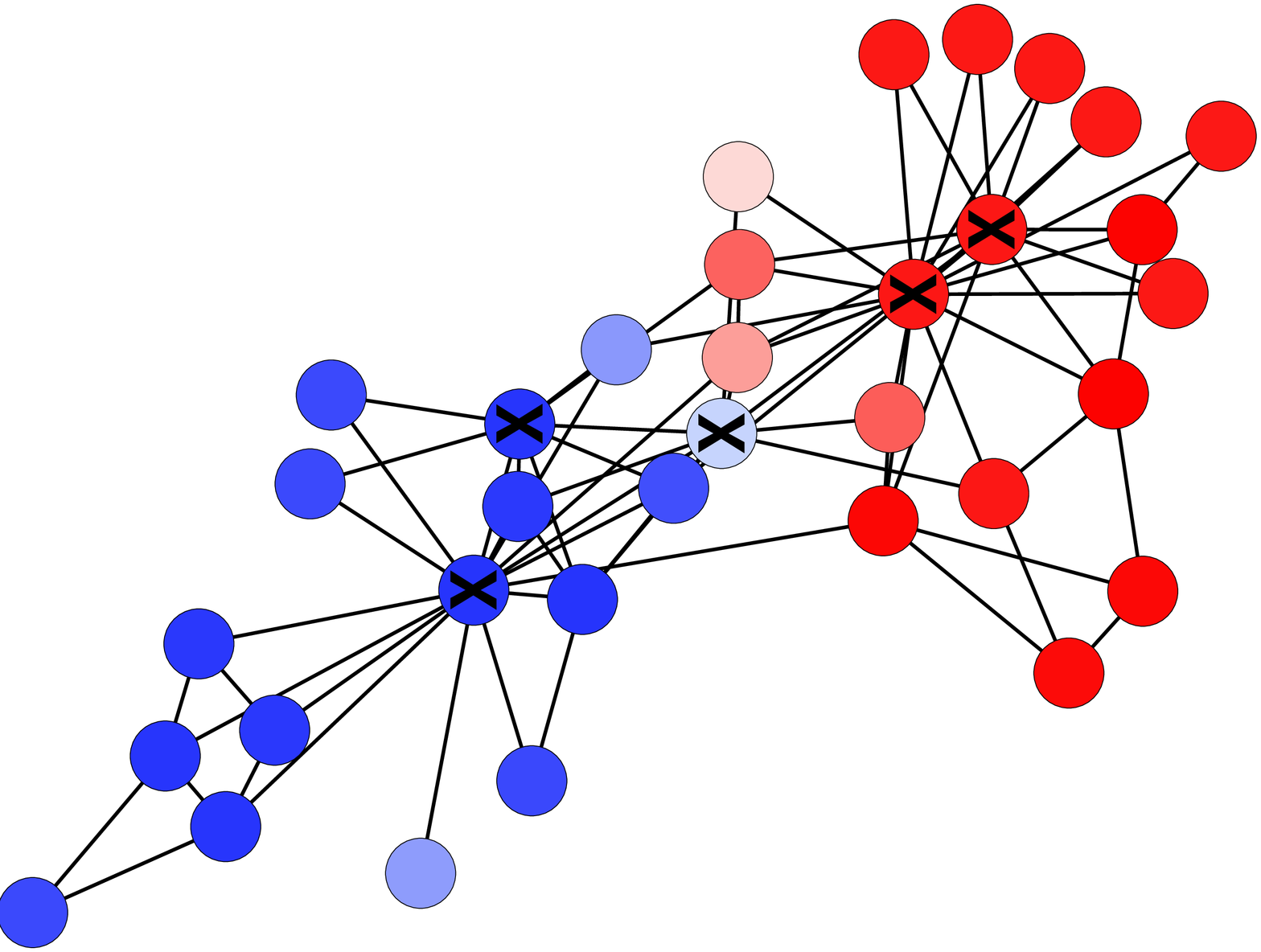}
   \includegraphics[width=0.495\linewidth]{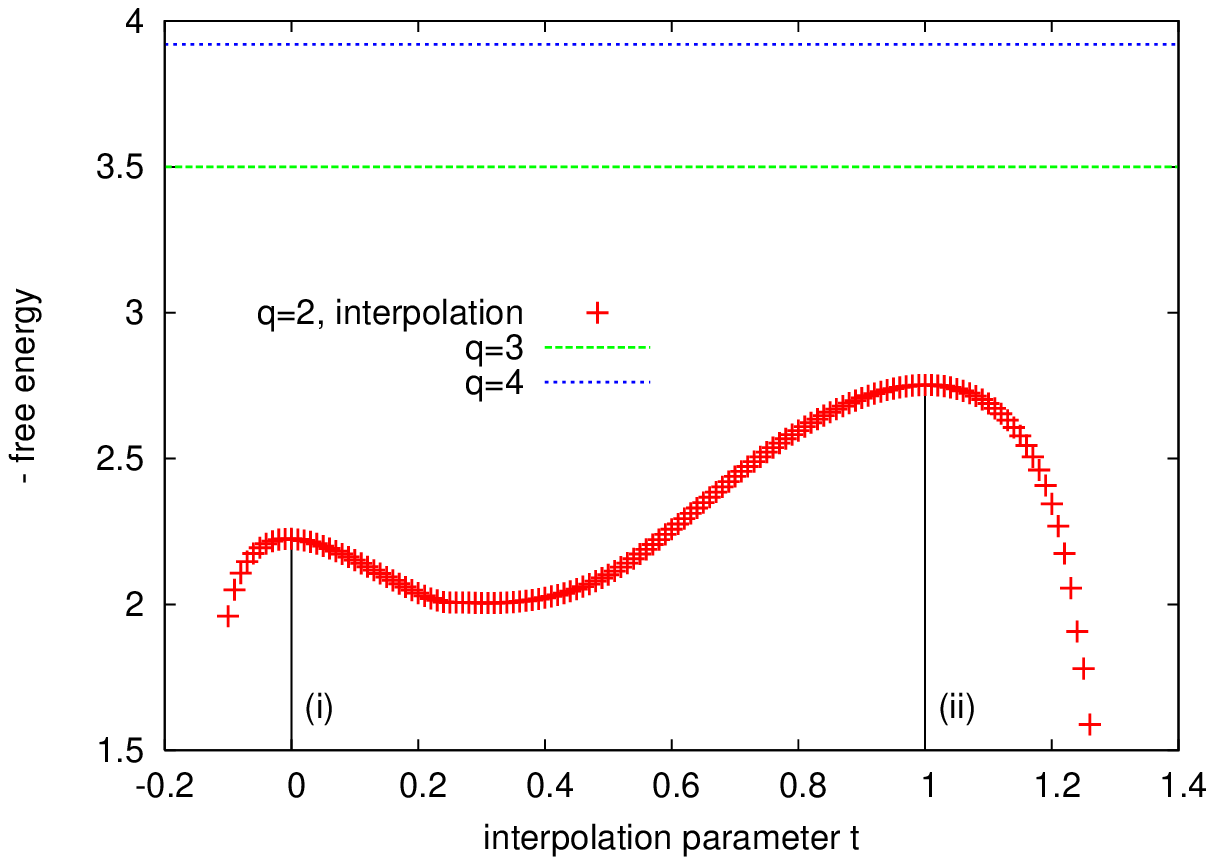}
  \caption{\label{fig_6} (color online): On the left: the partitioning of Zachary's karate club found by our inference algorithm using the first fixed point, $(i)$ in~\eqref{bp-fixedpoints}. The colors indicate the two groups found by starting with an assortative initial condition, i.e., where $c_{11}, c_{22} > c_{12}$.  The shades represent the marginal probabilities: a white node belongs to both groups with equal probability, whereas a node that is solid red or solid blue belongs to the corresponding group with probability $1$.  Most of the nodes are strongly biased.  
The $\times$s show the five nodes that are grouped together by the second fixed point, $(ii)$ in~\eqref{bp-fixedpoints}, which divides the nodes into high-degree and low-degree groups rather than into the two factions. On the right: the negative free energy for parameters interpolating between the two fixed points, with $(i)$ at $t=0$ and $(ii)$ at $t=1$.  
The two fixed points are local maxima, and each one has a basin of attraction in the learning algorithm.  As noted in~\cite{KarrerNewman10}, the high-degree/low-degree fixed point actually has lower free energy, and hence a higher likelihood, in the space of block models with $q=2$.  The horizontal lines show the largest values of the likelihood that we obtained from using more than two groups.  Unlike in Fig.~\ref{fig_5}, the likelihood continues to increase when more groups are allowed.  
This is due both to finite-size effects and to the fact that the network is not, in fact, generated by the block model: 
in particular, the nodes in each faction have a highly inhomogeneous degree distribution.}  
\end{figure}

Fig.~\ref{fig_6} shows the marginalized group assignments for the division into two groups corresponding to these two fixed points.  Fixed point $(i)$ corresponds to the actual division into two factions, and $c_{ab}^{(i)}$ has assortative structure, with larger affinities on the diagonal.  In contrast, fixed point $(ii)$ divides the nodes according to their degree, placing high-degree nodes in one group, including both the president and the instructor, and the low-degree nodes in the other group.  Of course, this second division is not wrong; rather, it focuses on a different kind of classification, into ``leaders'' on the one hand and ``students/followers'' on the other. On the right side of Fig.~\ref{fig_6} we plot the negative free energy~(\ref{Bethe_fe}) achieved by interpolating between the two fixed points according to a parameter $t$, with $c_{ab}(t) = (1-t) c_{ab}^{(i)} + t c_{ab}^{(ii)}$ and similarly for $n_a$.  
We see that the two fixed points correspond to two local maxima, the second $(ii)$ being the global one.  Thus if we assume that the network was generated by a block model with $q=2$, the second fixed point is the more likely division.

As recently pointed out in~\cite{KarrerNewman10}, the block model we study in this paper does not fit Zachary's network particularly well.  This is because the nodes in each faction are not equivalent to each other.  In particular, the ``hubs'' or ``leaders'' of each faction have significantly higher degrees than the other nodes do; in our block model this is unlikely, since the degree distribution within each group is Poisson.  The authors of~\cite{KarrerNewman10} show that we can obtain a better classification, in the sense of being closer to the two factions, using a \emph{degree-corrected} block model that takes this inhomogeneous degree distribution into account.  Happily, as we will discuss in future work, our BP approach and learning algorithm generalizes easily to these degree corrected block models.  Under the degree-corrected block model the factional division $(i)$ does indeed become the most likely one.

\begin{figure}[!ht]
   \includegraphics[width=0.495\linewidth]{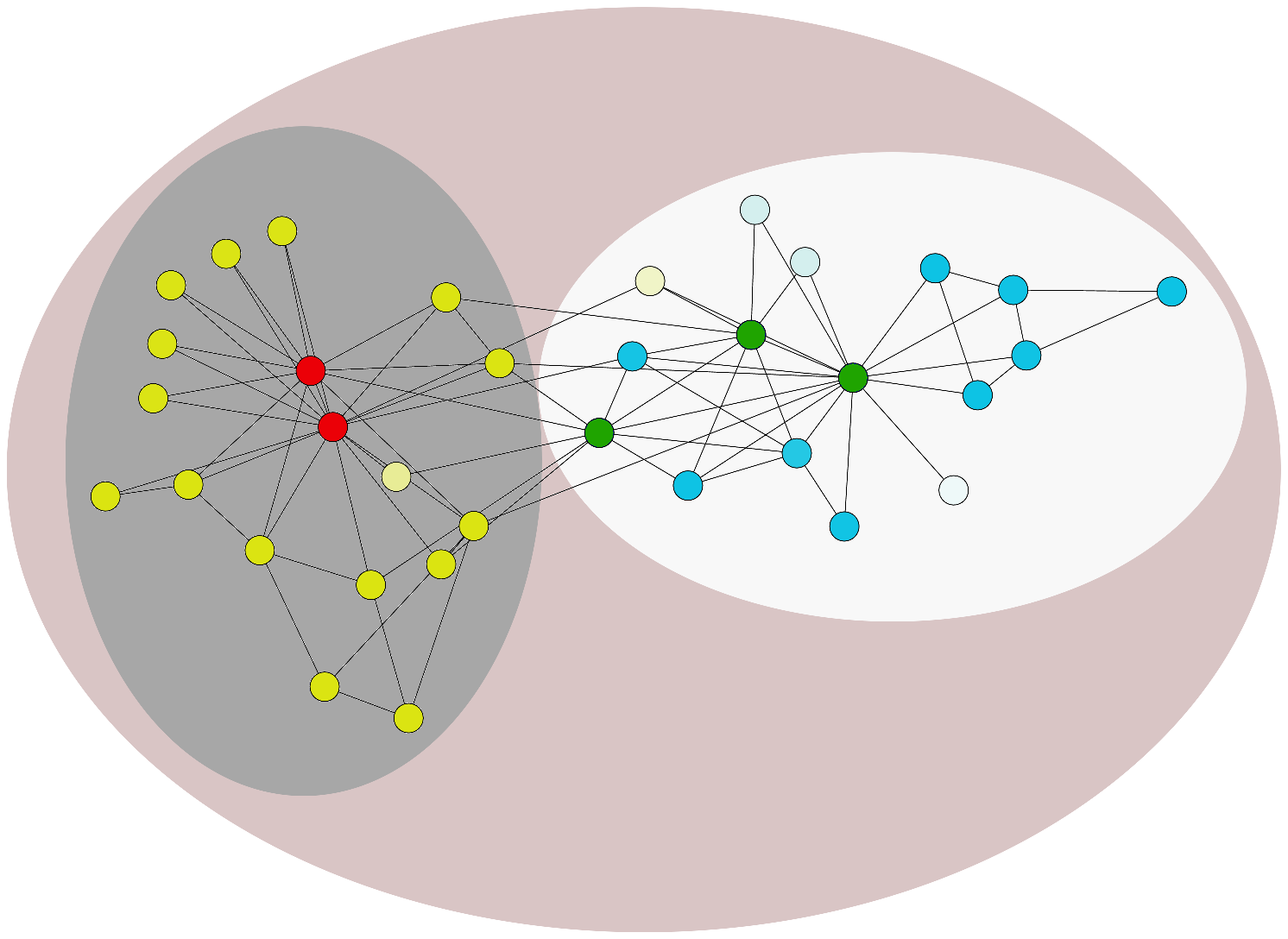}
   \includegraphics[width=0.495\linewidth]{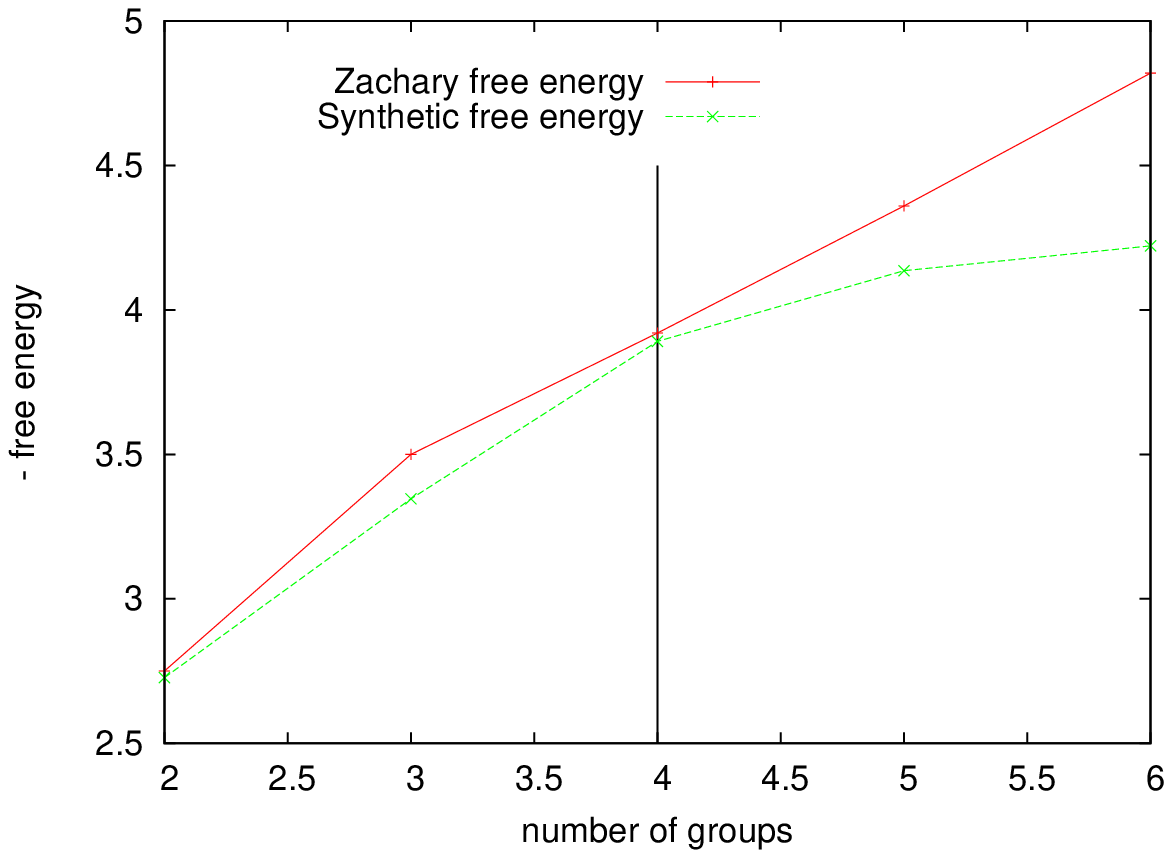}
  \caption{\label{fig_7} (color online): On the left, the marginalized group assignment for Zachary's karate club with $q=4$.  We again use the shade to indicate how strongly biased the marginal is, up to white if it is $1/q$.  Almost all nodes are strongly biased.  The white and dark grey regions correspond to the two factions, and within each group we have high- and low-degree nodes.  Thus our algorithm finds a partition that divides nodes according to both their faction and their degree.  
On the right, we compare the negative free energy (i.e., the likelihood) as a function of the number of groups for Zachary's network and a synthetic network with the same parameters as the $q=4$ fixed point.  The free energy levels off at $q=4$ for the synthetic network, but not as sharply as it does in the thermodynamic limit (see Fig.~\ref{fig_5}).  For Zachary's network, the free energy continues to improve as $q$ increases, due to further inhomogeneities within the groups.
}
\end{figure}

As discussed in Section~\ref{fact_learning}, our methods allow us to learn the correct number of groups $q^*$ for large networks generated from the block model by comparing the best free energy achieved for different $q$.  In the thermodynamic limit, the free energy becomes constant when $q \ge q^*$ as shown in Fig.~\ref{fig_5}.  However, for real networks the situation is less simple.  
As shown on the right of Fig.~\ref{fig_6}, for Zachary's network our algorithm finds fixed points with decreasing free energy (i.e., increasing likelihood) as $q$ increases.  On the left side of Fig.~\ref{fig_7}, we show the marginalized group assignment we found for $q=4$.  As we would expect given the fact that each faction contains both high- and low-degree nodes, the four-group classification separates nodes according to both their faction and their degree, dividing each faction into leaders and followers.  As $q$ increases further, so does the likelihood, due to the fact that there are further inhomogeneities within the groups, i.e., further deviations from the block model.  Continuing to subdivide the groups leads to a hierarchy of groups and subgroups as in~\cite{ClausetMoore08}, for instance separating peripheral nodes in each faction from those that are directly connected to the hubs.

However, even for networks generated by the block model, there are finite-size effects that make learning the number of groups difficult.  To separate these finite-size effects from the inhomogeneities in Zachary's network, we used the block model to generate synthetic networks, using the parameters that our algorithm learned for Zachary's network with $q=4$.  On the right side of Fig.~\ref{fig_7} we show the negative free energy obtained by our algorithm on the real and synthetic networks for various values of $q$.  For the synthetic networks, the likelihood levels off to some extent when $q \ge q^*$, but does not become constant.  Thus these finite-size effects explain some, but not all, of the likelihood increase observed for the real network as $q$ increases.

\subsection{A network of political books}

\begin{figure}[!ht]
   \includegraphics[width=0.6\linewidth, bb= 18 271 577 572,clip]{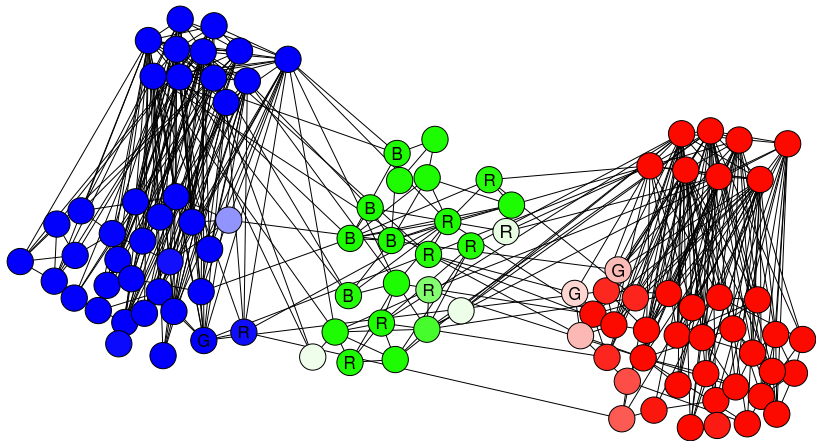}
  \caption{\label{fig_8} (color online): 
The marginalized group assignment with $q=3$ for the political book network, with liberal, neutral, and conservative books labeled red, green, and blue respectively.  The letters R, G, and B indicate the original labels where they disagree with our algorithm.  As for Zachary's network, the likelihood increases as we increase the number of groups.  For $q=5$, running our algorithm with $q=5$ subdivides the red and blue groups into subcommunities as shown.}
\end{figure}

The other real network we discuss here is a network of political books sold on Amazon compiled by V. Krebs (unpublished). These books are labeled according to three groups, liberal, neutral, or conservative.  Edges represent co-purchasing of books by the same customer as reported by Amazon.  Running our learning BP algorithm with $q=3$ yields a group assignment with an overlap of $Q=0.74$.  The parameters learned by our algorithm, and the most-likely parameters given the original labeling, are respectively
\bea
  n^{(i)} =
  \left(
   \begin{array}{c}
     0.24 \\
     0.39 \\
     0.37
   \end{array} 
  \right) \; , \;\quad
  c^{(i)} = 
   \left(
   \begin{array}{ccc}
     18.0 & 2.7 & 2.1 \\
     2.7 & 21.2 & 0.15 \\
     2.1 & 0.15 & 22.6
   \end{array} 
  \right) 
 \\
  n_{\rm actual} =
  \left(
   \begin{array}{c}
     0.13 \\
     0.46 \\
     0.41
   \end{array} 
  \right) \; , \; \quad
  c_{\rm actual} = 
   \left(
   \begin{array}{ccc}
     12.1 & 5.6 & 4.5\\
     5.6 & 17 & 0.6 \\
     4.5 & 0.6 & 20
   \end{array} 
  \right) \, .
\eea
The marginalized group assignment for $q=3$ our algorithm finds using the learned parameters $c_{ab}^{(i)}$, $n_a^{(i)}$ is shown in Fig.~\ref{fig_8}.  The letters correspond to the actual group assignment for nodes where our assignment disagrees with Krebs' labels.  The groups corresponding to liberal and conservative books agree very well with his labels, and are well-separated with very few links between them.  However, the middle group found by our algorithm is a mixture of the three types of books. It could be that the original labels of these books are misleading, or that these particular books appeal to customers who buy books from all three parts of the political spectrum.

We also ran our algorithm with $q \ne 3$ groups.  Using $q=2$ distributes the neutral books in the two larger groups.  For $q=5$, the liberal and conservative groups each break into two subgroups as shown in Fig.~\ref{fig_8}, which consist mainly of high-degree and low-degree nodes within each of these groups. This structure can be observed in the corresponding affinity matrix
\be
  n^{(ii)} =
  \left(
   \begin{array}{c}
     0.23 \\
     0.32 \\
     0.08 \\
     0.27 \\
     0.1
   \end{array} 
  \right) \; , \; \quad
  c^{(ii)} = 
   \left(
   \begin{array}{ccccc}
     21.4 & 1.54 & 6.1 & 0.64 & 6.1 \\
     1.54 & 8.3 & 38.4 & 0.3 & 0 \\
     6.1 & 38.4 & 65.5 & 0 & 0 \\
     0.64 & 0.3 & 0 & 8.1 & 33.1 \\
     6.1 & 0 & 0 & 33.1 & 62.6
   \end{array} 
  \right) \, ,
\ee
where group 1 consists of the neutral books, groups 2 and 3 are the low- and high-degree liberal books, and 4 and 5 are the low- and high-degree conservative books. 
However, as for Zachary's karate club, the likelihood keeps increasing for larger $q$, suggesting a hierarchy of groups or subgroups, or simply that the block model is not the right generative model for this network.

\section{Conclusions}

We analyzed the thermodynamic properties of networks generated by the stochastic block model, focusing on the questions of how to optimally infer the original group assignment, and learn the parameters of the model, from the topology of the generated graph. Using the cavity method we provided an asymptotically exact answer to these questions, describing the phase diagram and transitions between phases where inference is easy, possible but exponentially hard, and impossible.  These transitions are closely related to known phase transitions in the mean field theory of spin glasses.  Remaining open questions include an analysis of finite size effects, and mathematically rigorous proofs of our results. 

In the easy phase, our analysis leads to a belief propagation (BP) algorithm that infers the group assignment optimally, i.e., that maximizes the overlap with the original assignment, and learns the underlying parameters, including the correct number of groups in the thermodynamic limit.  This algorithm is highly scalable, with a running time that is linear in the size of the network.  While MCMC sampling also runs in linear time, BP is considerably faster at providing the marginal probabilities, which also give a measure of how strongly each node belongs to its group.  Our methods can detect more general types of functional communities than many other methods of community detection algorithms, and also provide measures of significance of the community structure, letting us distinguish purely random graphs from those with modular structure.  

While many real-world networks are not well modeled by the type of stochastic block model we study here, our analysis and our BP learning algorithm easily generalize to any generative model where the likelihood function (\ref{Pgraph}) can be written as a product of local terms, such as the degree-corrected block models of~\cite{KarrerNewman10}.  We will discuss these generalizations in future work.

\bibliography{myentries}

\begin{thebibliography}{46}
\expandafter\ifx\csname natexlab\endcsname\relax\def\natexlab#1{#1}\fi
\expandafter\ifx\csname bibnamefont\endcsname\relax
  \def\bibnamefont#1{#1}\fi
\expandafter\ifx\csname bibfnamefont\endcsname\relax
  \def\bibfnamefont#1{#1}\fi
\expandafter\ifx\csname citenamefont\endcsname\relax
  \def\citenamefont#1{#1}\fi
\expandafter\ifx\csname url\endcsname\relax
  \def\url#1{\texttt{#1}}\fi
\expandafter\ifx\csname urlprefix\endcsname\relax\def\urlprefix{URL }\fi
\providecommand{\bibinfo}[2]{#2}
\providecommand{\eprint}[2][]{\url{#2}}

\bibitem[{\citenamefont{Fortunato}(2010)}]{Fortunato10}
\bibinfo{author}{\bibfnamefont{S.}~\bibnamefont{Fortunato}},
  \bibinfo{journal}{Physics Reports} \textbf{\bibinfo{volume}{486}},
  \bibinfo{pages}{75} (\bibinfo{year}{2010}).

\bibitem[{\citenamefont{Erd{\H{o}}s and R{\'e}nyi}(1959)}]{ErdosRenyi59}
\bibinfo{author}{\bibfnamefont{P.}~\bibnamefont{Erd{\H{o}}s}} \bibnamefont{and}
  \bibinfo{author}{\bibfnamefont{A.}~\bibnamefont{R{\'e}nyi}},
  \bibinfo{journal}{Publ. Math. Debrecen} \textbf{\bibinfo{volume}{6}},
  \bibinfo{pages}{290} (\bibinfo{year}{1959}).

\bibitem[{\citenamefont{Decelle et~al.}(2011)\citenamefont{Decelle, Krzakala,
  Moore, and Zdeborov\'a}}]{DecelleKrzakala11}
\bibinfo{author}{\bibfnamefont{A.}~\bibnamefont{Decelle}},
  \bibinfo{author}{\bibfnamefont{F.}~\bibnamefont{Krzakala}},
  \bibinfo{author}{\bibfnamefont{C.}~\bibnamefont{Moore}}, \bibnamefont{and}
  \bibinfo{author}{\bibfnamefont{L.}~\bibnamefont{Zdeborov\'a}},
  \bibinfo{journal}{Phys. Rev. Lett.} \textbf{\bibinfo{volume}{107}},
  \bibinfo{pages}{065701} (\bibinfo{year}{2011}).

\bibitem[{\citenamefont{M\'ezard and Montanari}(2009)}]{MezardMontanari07}
\bibinfo{author}{\bibfnamefont{M.}~\bibnamefont{M\'ezard}} \bibnamefont{and}
  \bibinfo{author}{\bibfnamefont{A.}~\bibnamefont{Montanari}},
  \emph{\bibinfo{title}{Physics, Information, Computation}}
  (\bibinfo{publisher}{Oxford Press}, \bibinfo{address}{Oxford},
  \bibinfo{year}{2009}).

\bibitem[{\citenamefont{Yedidia et~al.}(2003)\citenamefont{Yedidia, Freeman,
  and Weiss}}]{YedidiaFreeman03}
\bibinfo{author}{\bibfnamefont{J.}~\bibnamefont{Yedidia}},
  \bibinfo{author}{\bibfnamefont{W.}~\bibnamefont{Freeman}}, \bibnamefont{and}
  \bibinfo{author}{\bibfnamefont{Y.}~\bibnamefont{Weiss}}, in
  \emph{\bibinfo{booktitle}{Exploring Artificial Intelligence in the New
  Millennium}} (\bibinfo{publisher}{Science \& Technology Books},
  \bibinfo{year}{2003}), pp. \bibinfo{pages}{239--236}.

\bibitem[{\citenamefont{Clauset et~al.}(2008)\citenamefont{Clauset, Moore, and
  Newman}}]{ClausetMoore08}
\bibinfo{author}{\bibfnamefont{A.}~\bibnamefont{Clauset}},
  \bibinfo{author}{\bibfnamefont{C.}~\bibnamefont{Moore}}, \bibnamefont{and}
  \bibinfo{author}{\bibfnamefont{M.}~\bibnamefont{Newman}},
  \bibinfo{journal}{Nature} \textbf{\bibinfo{volume}{453}}, \bibinfo{pages}{98
  } (\bibinfo{year}{2008}).

\bibitem[{\citenamefont{Airoldi et~al.}(2008)\citenamefont{Airoldi, Blei,
  Fienberg, and Xing}}]{airoldi}
\bibinfo{author}{\bibfnamefont{E.~M.} \bibnamefont{Airoldi}},
  \bibinfo{author}{\bibfnamefont{D.~M.} \bibnamefont{Blei}},
  \bibinfo{author}{\bibfnamefont{S.~E.} \bibnamefont{Fienberg}},
  \bibnamefont{and} \bibinfo{author}{\bibfnamefont{E.~P.} \bibnamefont{Xing}},
  \bibinfo{journal}{J. Machine Learning Research} \textbf{\bibinfo{volume}{9}},
  \bibinfo{pages}{1981} (\bibinfo{year}{2008}).

\bibitem[{\citenamefont{Karrer and Newman}(2011)}]{KarrerNewman10}
\bibinfo{author}{\bibfnamefont{B.}~\bibnamefont{Karrer}} \bibnamefont{and}
  \bibinfo{author}{\bibfnamefont{M.~E.~J.} \bibnamefont{Newman}},
  \bibinfo{journal}{Phys. Rev. E} \textbf{\bibinfo{volume}{83}},
  \bibinfo{pages}{016107} (\bibinfo{year}{2011}).

\bibitem[{\citenamefont{Newman and Girvan}(2004)}]{NewmanGirvan04}
\bibinfo{author}{\bibfnamefont{M.~E.~J.} \bibnamefont{Newman}}
  \bibnamefont{and} \bibinfo{author}{\bibfnamefont{M.}~\bibnamefont{Girvan}},
  \bibinfo{journal}{Phys. Rev. E} \textbf{\bibinfo{volume}{69}},
  \bibinfo{pages}{026113} (\bibinfo{year}{2004}).

\bibitem[{\citenamefont{Dyer and Frieze}(1989)}]{DyerFrieze89}
\bibinfo{author}{\bibfnamefont{M.~E.} \bibnamefont{Dyer}} \bibnamefont{and}
  \bibinfo{author}{\bibfnamefont{A.~M.} \bibnamefont{Frieze}},
  \bibinfo{journal}{Journal of Algorithms} \textbf{\bibinfo{volume}{10}},
  \bibinfo{pages}{451} (\bibinfo{year}{1989}).

\bibitem[{\citenamefont{Condon and Karp}(2001)}]{CondonKarp01}
\bibinfo{author}{\bibfnamefont{A.}~\bibnamefont{Condon}} \bibnamefont{and}
  \bibinfo{author}{\bibfnamefont{R.~M.} \bibnamefont{Karp}},
  \bibinfo{journal}{Random Struct. Algor.} \textbf{\bibinfo{volume}{18}},
  \bibinfo{pages}{116} (\bibinfo{year}{2001}).

\bibitem[{\citenamefont{Krivelevich and
  Vilenchik}(2006)}]{KrivelevichVilenchik06}
\bibinfo{author}{\bibfnamefont{M.}~\bibnamefont{Krivelevich}} \bibnamefont{and}
  \bibinfo{author}{\bibfnamefont{D.}~\bibnamefont{Vilenchik}}, in
  \emph{\bibinfo{booktitle}{Proceedings of the Third Workshop on Analytic
  Algorithmics and Combinatorics (ANALCO)}} (\bibinfo{year}{2006}), pp.
  \bibinfo{pages}{211--221}.

\bibitem[{\citenamefont{Krzakala and Zdeborov\'a}(2009)}]{KrzakalaZdeborova09}
\bibinfo{author}{\bibfnamefont{F.}~\bibnamefont{Krzakala}} \bibnamefont{and}
  \bibinfo{author}{\bibfnamefont{L.}~\bibnamefont{Zdeborov\'a}},
  \bibinfo{journal}{Phys. Rev. Lett.} \textbf{\bibinfo{volume}{102}},
  \bibinfo{pages}{238701} (\bibinfo{year}{2009}).

\bibitem[{\citenamefont{Bickel and Chen}(2009)}]{BickelChen09}
\bibinfo{author}{\bibfnamefont{P.~J.} \bibnamefont{Bickel}} \bibnamefont{and}
  \bibinfo{author}{\bibfnamefont{A.}~\bibnamefont{Chen}},
  \bibinfo{journal}{Proc. natl. Acad. Sci USA} \textbf{\bibinfo{volume}{106}}
  (\bibinfo{year}{2009}).

\bibitem[{\citenamefont{Iba}(1999)}]{Iba99}
\bibinfo{author}{\bibfnamefont{Y.}~\bibnamefont{Iba}},
  \bibinfo{journal}{Journal of Physics A: Mathematical and General}
  \textbf{\bibinfo{volume}{32}}, \bibinfo{pages}{3875} (\bibinfo{year}{1999}).

\bibitem[{\citenamefont{Dempster et~al.}(1977)\citenamefont{Dempster, Laird,
  and Rubin}}]{DempsterLaird77}
\bibinfo{author}{\bibfnamefont{A.}~\bibnamefont{Dempster}},
  \bibinfo{author}{\bibfnamefont{N.}~\bibnamefont{Laird}}, \bibnamefont{and}
  \bibinfo{author}{\bibfnamefont{D.}~\bibnamefont{Rubin}},
  \bibinfo{journal}{Journal of the Royal Statistical Society}
  \textbf{\bibinfo{volume}{39}}, \bibinfo{pages}{1–38}
  (\bibinfo{year}{1977}).

\bibitem[{\citenamefont{Good et~al.}(2010)\citenamefont{Good, de~Montjoye, and
  Clauset}}]{GoodMontjoye10}
\bibinfo{author}{\bibfnamefont{B.~H.} \bibnamefont{Good}},
  \bibinfo{author}{\bibfnamefont{Y.-A.} \bibnamefont{de~Montjoye}},
  \bibnamefont{and} \bibinfo{author}{\bibfnamefont{A.}~\bibnamefont{Clauset}},
  \bibinfo{journal}{Physical Review E} \textbf{\bibinfo{volume}{81}},
  \bibinfo{pages}{046106} (\bibinfo{year}{2010}).

\bibitem[{\citenamefont{Newman and Leicht}(2007)}]{NewmanLeicht07}
\bibinfo{author}{\bibfnamefont{M.~E.~J.} \bibnamefont{Newman}}
  \bibnamefont{and} \bibinfo{author}{\bibfnamefont{E.~A.}
  \bibnamefont{Leicht}}, \bibinfo{journal}{Proc. Natl. Acad. Sci. USA}
  \textbf{\bibinfo{volume}{104}}, \bibinfo{pages}{9564} (\bibinfo{year}{2007}).

\bibitem[{\citenamefont{Ball et~al.}(2011)\citenamefont{Ball, Karrer, and
  Newman}}]{BallKarrer11}
\bibinfo{author}{\bibfnamefont{B.}~\bibnamefont{Ball}},
  \bibinfo{author}{\bibfnamefont{B.}~\bibnamefont{Karrer}}, \bibnamefont{and}
  \bibinfo{author}{\bibfnamefont{M.~E.~J.} \bibnamefont{Newman}}
  (\bibinfo{year}{2011}), \bibinfo{note}{preprint arXiv:1104.3590}.

\bibitem[{\citenamefont{Hofman and Wiggins}(2008)}]{HofmanWiggins08}
\bibinfo{author}{\bibfnamefont{J.~M.} \bibnamefont{Hofman}} \bibnamefont{and}
  \bibinfo{author}{\bibfnamefont{C.~H.} \bibnamefont{Wiggins}},
  \bibinfo{journal}{Phys. Rev. Lett.} \textbf{\bibinfo{volume}{100}},
  \bibinfo{pages}{258701} (\bibinfo{year}{2008}).

\bibitem[{\citenamefont{Moore et~al.}(2011)\citenamefont{Moore, Yan, Zhu,
  Rouquier, and Lane}}]{MooreKDD}
\bibinfo{author}{\bibfnamefont{C.}~\bibnamefont{Moore}},
  \bibinfo{author}{\bibfnamefont{X.}~\bibnamefont{Yan}},
  \bibinfo{author}{\bibfnamefont{Y.}~\bibnamefont{Zhu}},
  \bibinfo{author}{\bibfnamefont{J.-B.} \bibnamefont{Rouquier}},
  \bibnamefont{and} \bibinfo{author}{\bibfnamefont{T.}~\bibnamefont{Lane}},
  \bibinfo{journal}{Proc. KDD}  (\bibinfo{year}{2011}).

\bibitem[{\citenamefont{M{\'e}zard and Parisi}(2001)}]{MezardParisi01}
\bibinfo{author}{\bibfnamefont{M.}~\bibnamefont{M{\'e}zard}} \bibnamefont{and}
  \bibinfo{author}{\bibfnamefont{G.}~\bibnamefont{Parisi}},
  \bibinfo{journal}{Eur. Phys. J. B} \textbf{\bibinfo{volume}{20}},
  \bibinfo{pages}{217} (\bibinfo{year}{2001}).

\bibitem[{\citenamefont{M\'ezard and Montanari}(2006)}]{MezardMontanari06}
\bibinfo{author}{\bibfnamefont{M.}~\bibnamefont{M\'ezard}} \bibnamefont{and}
  \bibinfo{author}{\bibfnamefont{A.}~\bibnamefont{Montanari}},
  \bibinfo{journal}{J. Stat. Phys.} \textbf{\bibinfo{volume}{124}},
  \bibinfo{pages}{1317} (\bibinfo{year}{2006}).

\bibitem[{\citenamefont{Zdeborov{\'a} and
  Krzakala}(2007)}]{ZdeborovaKrzakala07}
\bibinfo{author}{\bibfnamefont{L.}~\bibnamefont{Zdeborov{\'a}}}
  \bibnamefont{and} \bibinfo{author}{\bibfnamefont{F.}~\bibnamefont{Krzakala}},
  \bibinfo{journal}{Phys. Rev. E} \textbf{\bibinfo{volume}{76}},
  \bibinfo{pages}{031131} (\bibinfo{year}{2007}).

\bibitem[{\citenamefont{Krzakala et~al.}(2007)\citenamefont{Krzakala,
  Montanari, Ricci-Tersenghi, Semerjian, and
  Zdeborov{\'a}}}]{KrzakalaMontanari06}
\bibinfo{author}{\bibfnamefont{F.}~\bibnamefont{Krzakala}},
  \bibinfo{author}{\bibfnamefont{A.}~\bibnamefont{Montanari}},
  \bibinfo{author}{\bibfnamefont{F.}~\bibnamefont{Ricci-Tersenghi}},
  \bibinfo{author}{\bibfnamefont{G.}~\bibnamefont{Semerjian}},
  \bibnamefont{and}
  \bibinfo{author}{\bibfnamefont{L.}~\bibnamefont{Zdeborov{\'a}}},
  \bibinfo{journal}{Proc. Natl. Acad. Sci. U.S.A}
  \textbf{\bibinfo{volume}{104}}, \bibinfo{pages}{10318}
  (\bibinfo{year}{2007}).

\bibitem[{\citenamefont{Kauzmann}(1948)}]{Kauzmann48}
\bibinfo{author}{\bibfnamefont{W.}~\bibnamefont{Kauzmann}},
  \bibinfo{journal}{Chem. Rev.} \textbf{\bibinfo{volume}{43}},
  \bibinfo{pages}{219} (\bibinfo{year}{1948}).

\bibitem[{\citenamefont{Nishimori}(1993)}]{Nishimori93}
\bibinfo{author}{\bibfnamefont{H.}~\bibnamefont{Nishimori}},
  \bibinfo{journal}{J. Phys. Soc. Jpn.} \textbf{\bibinfo{volume}{62}},
  \bibinfo{pages}{2973} (\bibinfo{year}{1993}).

\bibitem[{\citenamefont{Sourlas}(1994)}]{Sourlas94}
\bibinfo{author}{\bibfnamefont{N.}~\bibnamefont{Sourlas}},
  \bibinfo{journal}{Europhys. Lett.} \textbf{\bibinfo{volume}{25}},
  \bibinfo{pages}{159} (\bibinfo{year}{1994}).

\bibitem[{\citenamefont{Nishimori}(2001)}]{NishimoriBook01}
\bibinfo{author}{\bibfnamefont{H.}~\bibnamefont{Nishimori}},
  \emph{\bibinfo{title}{Statistical Physics of Spin Glasses and Information
  Processing: An Introduction}} (\bibinfo{publisher}{Oxford University Press},
  \bibinfo{address}{Oxford, UK}, \bibinfo{year}{2001}).

\bibitem[{\citenamefont{Tanaka}(2002)}]{Tanaka02}
\bibinfo{author}{\bibfnamefont{K.}~\bibnamefont{Tanaka}}, \bibinfo{journal}{J.
  Phys. A: Math. Gen. 35} \textbf{\bibinfo{volume}{35}}, \bibinfo{pages}{R81}
  (\bibinfo{year}{2002}).

\bibitem[{\citenamefont{Hastings}(2006)}]{Hastings06}
\bibinfo{author}{\bibfnamefont{M.~B.} \bibnamefont{Hastings}},
  \bibinfo{journal}{Phys. Rev. E} \textbf{\bibinfo{volume}{74}},
  \bibinfo{pages}{035102} (\bibinfo{year}{2006}).

\bibitem[{\citenamefont{Lancichinetti and
  Fortunato}(2009)}]{LancichinettiFortunato09}
\bibinfo{author}{\bibfnamefont{A.}~\bibnamefont{Lancichinetti}}
  \bibnamefont{and}
  \bibinfo{author}{\bibfnamefont{S.}~\bibnamefont{Fortunato}},
  \bibinfo{journal}{Physical Review E} \textbf{\bibinfo{volume}{80}},
  \bibinfo{pages}{056117} (\bibinfo{year}{2009}).

\bibitem[{\citenamefont{\v{S}ulc and Zdeborov\'a}(2010)}]{SulcZdeborova10}
\bibinfo{author}{\bibfnamefont{P.}~\bibnamefont{\v{S}ulc}} \bibnamefont{and}
  \bibinfo{author}{\bibfnamefont{L.}~\bibnamefont{Zdeborov\'a}},
  \bibinfo{journal}{J. Phys. A: Math. Theor.} \textbf{\bibinfo{volume}{43}},
  \bibinfo{pages}{285003} (\bibinfo{year}{2010}).

\bibitem[{\citenamefont{M{\'e}zard et~al.}(1987)\citenamefont{M{\'e}zard,
  Parisi, and Virasoro}}]{MezardParisi87b}
\bibinfo{author}{\bibfnamefont{M.}~\bibnamefont{M{\'e}zard}},
  \bibinfo{author}{\bibfnamefont{G.}~\bibnamefont{Parisi}}, \bibnamefont{and}
  \bibinfo{author}{\bibfnamefont{M.~A.} \bibnamefont{Virasoro}},
  \emph{\bibinfo{title}{Spin-Glass Theory and Beyond}},
  vol.~\bibinfo{volume}{9} of \emph{\bibinfo{series}{Lecture Notes in Physics}}
  (\bibinfo{publisher}{World Scientific}, \bibinfo{address}{Singapore},
  \bibinfo{year}{1987}).

\bibitem[{\citenamefont{Franz et~al.}(2001)\citenamefont{Franz, M{\'e}zard,
  Ricci-Tersenghi, Weigt, and Zecchina}}]{FranzMezard01}
\bibinfo{author}{\bibfnamefont{S.}~\bibnamefont{Franz}},
  \bibinfo{author}{\bibfnamefont{M.}~\bibnamefont{M{\'e}zard}},
  \bibinfo{author}{\bibfnamefont{F.}~\bibnamefont{Ricci-Tersenghi}},
  \bibinfo{author}{\bibfnamefont{M.}~\bibnamefont{Weigt}}, \bibnamefont{and}
  \bibinfo{author}{\bibfnamefont{R.}~\bibnamefont{Zecchina}},
  \bibinfo{journal}{Europhys. Lett.} \textbf{\bibinfo{volume}{55}},
  \bibinfo{pages}{465} (\bibinfo{year}{2001}).

\bibitem[{\citenamefont{Montanari}(2008)}]{Montanari08}
\bibinfo{author}{\bibfnamefont{A.}~\bibnamefont{Montanari}},
  \bibinfo{journal}{European Transactions on Telecommunications}
  \textbf{\bibinfo{volume}{19}}, \bibinfo{pages}{385–403}
  (\bibinfo{year}{2008}).

\bibitem[{\citenamefont{de~Almeida and Thouless}(1978)}]{AlmeidaThouless78}
\bibinfo{author}{\bibfnamefont{J.~R.~L.} \bibnamefont{de~Almeida}}
  \bibnamefont{and} \bibinfo{author}{\bibfnamefont{D.~J.}
  \bibnamefont{Thouless}}, \bibinfo{journal}{J. Phys. A}
  \textbf{\bibinfo{volume}{11}}, \bibinfo{pages}{983} (\bibinfo{year}{1978}).

\bibitem[{\citenamefont{Kesten and
  Stigum}(1966{\natexlab{a}})}]{KestenStigum66}
\bibinfo{author}{\bibfnamefont{H.}~\bibnamefont{Kesten}} \bibnamefont{and}
  \bibinfo{author}{\bibfnamefont{B.~P.} \bibnamefont{Stigum}},
  \bibinfo{journal}{The Annals of Mathematical Statistics}
  \textbf{\bibinfo{volume}{37}}, \bibinfo{pages}{1463}
  (\bibinfo{year}{1966}{\natexlab{a}}).

\bibitem[{\citenamefont{Kesten and
  Stigum}(1966{\natexlab{b}})}]{KestenStigum66b}
\bibinfo{author}{\bibfnamefont{H.}~\bibnamefont{Kesten}} \bibnamefont{and}
  \bibinfo{author}{\bibfnamefont{B.~P.} \bibnamefont{Stigum}},
  \bibinfo{journal}{J. Math. Anal. Appl.} \textbf{\bibinfo{volume}{17}},
  \bibinfo{pages}{309} (\bibinfo{year}{1966}{\natexlab{b}}).

\bibitem[{\citenamefont{Janson and Mossel}(2004)}]{JansonMossel04}
\bibinfo{author}{\bibfnamefont{S.}~\bibnamefont{Janson}} \bibnamefont{and}
  \bibinfo{author}{\bibfnamefont{E.}~\bibnamefont{Mossel}},
  \bibinfo{journal}{Ann. Probab.} \textbf{\bibinfo{volume}{32}},
  \bibinfo{pages}{2630} (\bibinfo{year}{2004}).

\bibitem[{\citenamefont{Achlioptas and
  Coja-Oghlan}(2008)}]{AchlioptasCoja-Oghlan08}
\bibinfo{author}{\bibfnamefont{D.}~\bibnamefont{Achlioptas}} \bibnamefont{and}
  \bibinfo{author}{\bibfnamefont{A.}~\bibnamefont{Coja-Oghlan}}
  (\bibinfo{year}{2008}), p. \bibinfo{pages}{793}.

\bibitem[{\citenamefont{Zdeborov\'a and Krzakala}(2010)}]{ZdeborovaKrzakala10}
\bibinfo{author}{\bibfnamefont{L.}~\bibnamefont{Zdeborov\'a}} \bibnamefont{and}
  \bibinfo{author}{\bibfnamefont{F.}~\bibnamefont{Krzakala}},
  \bibinfo{journal}{Phys. Rev. B} \textbf{\bibinfo{volume}{81}},
  \bibinfo{pages}{224205} (\bibinfo{year}{2010}).

\bibitem[{\citenamefont{Reichardt and Leone}(2008)}]{ReichardtLeone08}
\bibinfo{author}{\bibfnamefont{J.}~\bibnamefont{Reichardt}} \bibnamefont{and}
  \bibinfo{author}{\bibfnamefont{M.}~\bibnamefont{Leone}},
  \bibinfo{journal}{Phys. Rev. Lett.} \textbf{\bibinfo{volume}{101}},
  \bibinfo{pages}{078701} (\bibinfo{year}{2008}).

\bibitem[{\citenamefont{Hu et~al.}(2011)\citenamefont{Hu, Ronhovde, and
  Nussinov}}]{HuRonhovde10}
\bibinfo{author}{\bibfnamefont{D.}~\bibnamefont{Hu}},
  \bibinfo{author}{\bibfnamefont{P.}~\bibnamefont{Ronhovde}}, \bibnamefont{and}
  \bibinfo{author}{\bibfnamefont{Z.}~\bibnamefont{Nussinov}}
  (\bibinfo{year}{2011}), \bibinfo{note}{arXiv:1008.2699v3}.

\bibitem[{\citenamefont{Zachary}(1977)}]{Zachary77}
\bibinfo{author}{\bibfnamefont{W.~W.} \bibnamefont{Zachary}},
  \bibinfo{journal}{Journal of Anthropological Research}
  \textbf{\bibinfo{volume}{33}}, \bibinfo{pages}{452} (\bibinfo{year}{1977}).

\bibitem[{\citenamefont{Lancichinetti et~al.}(2008)\citenamefont{Lancichinetti,
  Fortunato, and Radicchi}}]{LancichinettiFortunato08}
\bibinfo{author}{\bibfnamefont{A.}~\bibnamefont{Lancichinetti}},
  \bibinfo{author}{\bibfnamefont{S.}~\bibnamefont{Fortunato}},
  \bibnamefont{and} \bibinfo{author}{\bibfnamefont{F.}~\bibnamefont{Radicchi}},
  \bibinfo{journal}{Physical Review E} \textbf{\bibinfo{volume}{78}},
  \bibinfo{pages}{046110} (\bibinfo{year}{2008}).

\end{thebibliography}

\end{document}